\documentclass[fleqn,usenatbib]{mnras}

\usepackage{txfonts}

\usepackage[T1]{fontenc}
\usepackage{ae,aecompl}

\usepackage{graphicx}

\usepackage{silence}
\usepackage{caption}
\usepackage{subcaption}
\usepackage{xcolor}
\usepackage{soul} 
\usepackage{psfrag}

\title[Detection and analysis of cosmological filaments]{Detection and analysis of cluster-cluster filaments} 
\author[Pereyra et al.]{Luis A. Pereyra$^{1}$\thanks{E-mail:luis.pereyra@unc.edu.ar}, Mario A. Sgr\'o$^{1,2}$, Manuel E. Merch\'an$^{1,2}$,\newauthor Federico A. Stasyszyn$^{1,2}$ and Dante J. Paz$^{1,2}$\\
$^{1}$ Instituto de Astrof\'isica Te\'orica y Experimental, CONICET - UNC, Laprida 854, C\'ordoba, Argentina\\
$^{2}$ Observatorio Astron\'omico de C\'ordoba, Universidad Nacional de C\'ordoba, Laprida 854, X5000BGR, C\'ordoba, Argentina.
}

\begin{document}
\label{firstpage}
\pagerange{\pageref{firstpage}--\pageref{lastpage}}
\maketitle

\begin{abstract}

In this work, we identify and analyse the properties of cluster-cluster filaments within a cosmological simulation assuming that they are structures connecting maxima of the density field defined by dark matter halos with masses $M \, \ge 10^{14}\, h^{-1} \mathrm{M_{\odot}}$.
To extract these filaments we develop an identification algorithm
based on two standard tools: the Minimal Spanning Tree (MST) and the Friends of Friends (FoF) algorithm. Focusing our analysis on the densest dark matter filaments, we found that the radial density profile, at scales around $1\, h^{-1} \mathrm{Mpc}$, approximately follow a power-law function with index -2.
Without making any assumption about the velocity field, our algorithm finds that the saddle point arises as a natural characteristic of the filamentary structure. In addition, its location along the filament
depends on the masses of the halos at the filament ends. We also found that the infall velocities follow a cross-pattern near the saddle point, being perpendicular to the filament spine when approaching from low-density regions, and parallel away from the saddle point towards the ends of the filament.
Following theoretical prescriptions, we estimate the linear density from the transverse velocity dispersion, finding a good correspondence with the measured mass per unit length of our filaments.
Our results can be applied to observational samples of filaments in order to link the saddle point location and the mass per unit length
with measurements obtained from observations such as cluster masses and the velocity dispersion of galaxies.

\end{abstract}

\begin{keywords}
methods: numerical, methods: statistical, large-scale structure of the Universe - Numerical Simulations
\end{keywords}

\section{Introduction}

The distribution of matter in the Universe 
at large scale forms a complex pattern known as
the cosmic web \citep{Bond1996}.
It is composed of dense compact clusters, elongated filaments, sheet-like walls and underdense regions called voids.
The Universe is dominated by clusters and filaments in terms of mass and by voids in terms of volume. The densest structures in the cosmic network beyond the halos are the filaments \citep{VanHaarlem1993, Colberg1999}.

The study and characterisation of filaments are interesting for different reasons. The cosmological simulations reveal the role of filaments over their environment. 
Several authors confirm the influence of these structures on the alignment of the angular momentum and the shape of dark matter halos with their large-scale surroundings
\citep{Altay2006, Hahn2007a, Hahn2007b, Paz2008, AragonCalvo2007b, Zhang2009, Paz2011, 
Libeskind2013, AragonCalvo2014, GaneshaiahVeena2018, GaneshaiahVeena2019}.
Studies based on observed galaxy catalogues show evidence of correlations between the orientation and/or
shape of the galaxies and the spine of the nearby filaments
\citep{Jones2010, Tempel2013a, Tempel2013b, Zhang2013, Zhang2015, Chen2019, Kraljic2019, Welker2020}.
Besides this alignment effect, the analysis of galaxy properties shows that the fraction of passive galaxies increases toward the spine of filaments both in galaxy catalogues and in hydrodynamic cosmological simulations.
\citep[e.g.][]{Alpaslan2016, Martinez2016, Malavasi2017, Chen2017, Kraljic2018, Laigle2018, Salerno2019}.
Recently, \citet{Rost2020} compare three different catalogues of filaments identified in the SDSS DR12, finding that the characteristics of galaxies populating filaments depends on the method employed to extract the filamentary structure. They also show that the shorter filaments are more populated by red galaxies and have better-defined overdensity profiles than their longer counterpart. Moreover, it has been shown that the filamentary structure influences the evolution of galaxy groups and their central galaxies.
\citep{Poudel2017, DarraghFord2019}.
In this sense, \cite{Guo2015} found that galaxy groups in filaments have more satellites than those residing outside of filamentary environments.
In a likewise manner 
other works \citep[see e.g.][]{Welker2017, Welker2018, Lee2018, GaneshaiahVeena2018, Tempel2015}
explored the alignment of satellite galaxies with respect to filaments, finding that the satellites tend to be aligned with them.
They suggest that the alignment signal may be a consequence of how satellites accrete matter via streams along the direction of the filaments.

The works mentioned above show that the filamentary network plays an important role in the systems that compose it. 
Consequently, the study of the filament intrinsic properties is important to understand their connection to the formation of dark matter halos and galaxies. Several works have focused on the analysis of filament properties in both numerical simulations and large catalogues of galaxies. Some of them have been interested in their baryonic component \citep[e.g.][]{Dolag2006, Gheller2015}, while other studies estimate its theoretical abundance \citep{Shen2006} or its geometric properties such as length, thickness or density profiles \citep{Cautun2014}.
A large number of studies in the literature confirm that the filamentary structure is closely affiliated to the clusters \citep[e.g.][]{Colberg2005,Gonzalez2010}.
Therefore, it should be expected that the intercluster bridges of matter dominate the infall of material into halos and galaxies located inside these prominent filaments, and subsequently influencing the formation process and properties of these structures.

In this context, filament identification plays a fundamental role in conducting these studies.
However, the detection of these objects represents a great challenge due to the complexity of their interconnected structures,
as well as the wide range of densities found in their distribution of matter.
Different techniques have been recently developed in the literature for this purpose 
e.g. 
the Multiscale Morphology Filter \citep{AragonCalvo2007a}, 
FINE \citep{Gonzalez2010},
Bisous model \citep{Stoica2010,Tempel2014a},
DisPerSE \citep{Sousbie2011a, Sousbie2011b}, 
NEXUS/NEXUS+ \citep{Cautun2013}, 
Adapted Minimal Spanning Tree \citep{Alpaslan2014}, 
SCMS \citep{Chen2015},
MultiStream Web Analysis \citep{Ramachandra2015}
and more recently T-ReX \citep{Bonnaire2020}.
A comprehensive comparison of different filament finding methods can be found in \citet{Libeskind2018}.
Most of these methods perform a reconstruction of the underlying density field and/or the velocity field that this induces, which makes them very effective in detecting different types of structures.
However, if the positions of halos or galaxies that are sparsely or irregularly distributed are used as tracers of the structure, it may not be possible to determine the eigenvalues of the Hessian matrix of the density field with sufficiently high precision \citep{Gonzalez2010}, thus avoiding the environment classification (cluster, filaments or walls) used by a hessian-based method. This case makes it necessary to use other independent techniques such as percolation algorithms or stochastic approaches.

Through this work, we implement an algorithm to extract the filamentary pattern in a cosmological simulation and analyse its statistical properties such as length, mass, and curliness distributions.
This procedure is based on the graph construction known as Minimal Spanning Tree (MST, hereafter) and the friend of friend algorithm (FoF, hereafter). 
We focus our study on the dynamical and structural properties of cluster-cluster filaments, for which we estimate their density profile, the location of the saddle point and the linear density, i.e. the density per unit length of these structures. 
The location of the saddle point is key to investigate the role of this characteristic in the evolution of the angular momentum of halos \citep[e.g.][]{Codis2015, Laigle2015}. However, the  determination of the position of the saddle point rely on indirect methods such as the one we present in this work. We explore the possibility of relating its location with a particular property of the filaments identified by our algorithm.
On the other hand, the linear density is an important property that can be compared with measurements obtained from observational data, as shown in the work of \citet{Eisenstein1997}, which found an analytical prescription to estimate this property trough the measure of the transverse velocity dispersion. 
Our investigations can provide an appropriate framework to compare with the results of upcoming galaxy surveys, such as HETDEX \citep{hetdex}, Euclid \citep{euclid}, DESI \citep{desi}, and LSST \citep{lsst}.

The organisation of this paper is as follows: 
in \S \ref{sec:metodo} we describe in detail the filaments identification method.
Afterwards, in \S \ref{sec:properties_filaments}, we apply this filament finder on a dark matter only cosmological simulation and expose global properties of the extracted structures.
In \S \ref{sec:dynamical_filaments} we study the dynamical and structural properties focusing our analysis on the stacked density field and the transverse velocity dispersion. Finally, we summarise our results and discuss future works in \S \ref{sec:summary_discussion}.

\section{
Filament detection
} \label{sec:metodo}

The evolution of the density field is well described at linear stages by the Zel'dovich approximation \citep{Zel1970}. 
In this context, filaments can be understood as highly asymmetrical structures consequence of the collapse of initial perturbations along two axes.
The structures that collapse along three axes correspond to nodes of the cosmic web traced by the filaments.

Based on these considerations,
we will adopt the basic approach to filaments embraced by several authors
\citep{Pimbblet2005,Colberg2005,Gonzalez2010,Martinez2016} according to whom the filaments are bridges of matter that connect high density peaks (cluster-cluster bridges). 
We want to be able to use any object as a tracer of the mass distribution (particles, dark matter halos, galaxies or clusters of galaxies), and to be capable of dealing with both numerical simulations and large catalogues of galaxies.
According to our approach, we would like to make sure that the nodes of the cosmic web are at the end of our filaments.

Taking into account
the aforementioned criteria,
we propose an algorithm based
on simple tools arranged in
the following five steps:

\begin{itemize}
    \item Apply a FoF algorithm to extract the tracers (if necessary) and the intermediate density regions in which the filaments are embedded.
    \item Use the MST technique to link the tracers restricting us to the intermediate density regions.
    \item Prune the tree to remove minor branches.
    \item Select filaments chopping the branches according to the mass of the tracers at their ends.
    \item Smooth the spine of each filament using B-spline technique.
\end{itemize}

Below we describe in detail each of these procedures applied to a cosmological simulation, however it can easily be extended to large catalogues of galaxies, as in \citet{Rost2020}.

\subsection{Friends-of-friends algorithm}
\label{sec:fof}
In this work we use dark matter halos as mass tracers, consequently, the first step of our method is to identify the halos in a simulation. For this purpose 
we use a FoF algorithm which puts together
all particles linked in pairs whenever their separation is less than a certain value (linking length). In our case, we select regions with an overdensity $\delta=200$ (with $\delta={\rho}/{\bar{\rho}}-1$ where $\rho$ is the local density and $\bar{\rho}$ is the mean density of the universe) which corresponds to the overdensity of a virialised halo and is associated to a standard linking length of $l_{1} = 0.17 \, n^{-1/3}$, where $n$ is the mean number density of particles in the simulation.

At this stage, the FoF algorithm is also used to extract
intermediate density regions that completely contain our
filaments. 
These regions will be used later to avoid filaments crossing low density regions, since we focus our analysis on cluster-cluster filaments and their exclusion reduce the computational cost. 
In particular, we chose an over-density of $\delta = 1$
(linking length $l_{2} = 0.79\, n^{-1/3}$), which is
approximately one order of magnitude less than the
typical over-density surrounding cosmological filaments
\citep{Cautun2013}.
Given that the overdensity of the wanted filaments is much higher, the value adopted for $l_{2}$ allows us to avoid low density regions without affecting our results. 

\subsection{Minimal Spanning Tree}
\label{sec:mst} 

In this work, we follow the definitions of \citet{Barrow1985} 
which were also previously used by other authors \citep[see e.g.][]{Colberg2007, Park2009, Alpaslan2014, Naidoo2020}.
Before describing the MST, it is important to introduce some key concepts of graph theory.
Given a set of objects, a graph $G$ is a collection of nodes and edges (straight lines that join nodes) relating those objects.
In particular, our method implements an undirect-weighted graph in which each edge has an associated weight ($\mathbf{w}$) and has no direction.
A spanning tree of $G$ is defined as a network connecting all the nodes on the graph.
The MST is the unique tree (if there are not two edges with equal weight) connecting all the nodes in $G$ so that the sum of weights is minimal and there are no closed loops on it.

A key property of the MST scheme is the degree $d$ of a given node, which is defined as the number of edges attached to that node. For example, bifurcation nodes of the tree have $d \ge 3$, meanwhile intermediate nodes have $d = 2$ and a node on a leaf of the tree has $d = 1$.
Finally, we define a $k$-branch as a path of $k$ edges connecting a node of degree $d = 1$ with another one of degree $d \ge 3$, being all the intermediate nodes of degree $d = 2$.

We will use the FoF halos as the nodes of the graph $G$. In order to define the edges of that graph, we construct a Voronoi tessellation of the simulation volume based on the positions of the nodes. For each FoF halo, we look for his Voronoi neighbours within the same intermediate density regions associated with the linking length $l_{2}$.
The Voronoi tessellation over nodes is computed using the public library VORO++ \citep{Rycroft2009}. 
Since the main physical process involved in the formation and evolution of the filaments is gravity, we decided to use a Newtonian 
edge weight ($\mathbf{w}$), defined as follows:

\begin{equation}
    \mathbf{w} = -\frac{M * m}{r^{2}}\,,
	\label{eq:weight}
\end{equation}
where $M$ and $m$ are the masses of the edges nodes (FoF halos) and $r$ is the distance between the centres of these objects.
Even though this weight is not derived from first principles, the criterion behind this choice is to emphasise the relative importance of the most massive and closest nodes.
At this point, it is worth mentioning that, strictly speaking, we have an MST for each intermediate density region, but for simplicity, we will continue our description as if there were only one MST since as we will see later, all filaments considered in this work inhabit a unique single region.

The MST of graph $G$ is constructed employing the Kruskal's algorithm \citep{Kruskal1956}. Briefly, we arrange the edges of $G$ by its weight in increasing order. After that, we sequentially select the lightest edge (the edge with lowest value of $w$) that do not produce closed loops on the tree.
Under the hypothesis that there are not two equal weights, which is a good assumption in our case, the selected edges represent the unique MST.

\begin{figure}
 \includegraphics[width=\columnwidth]{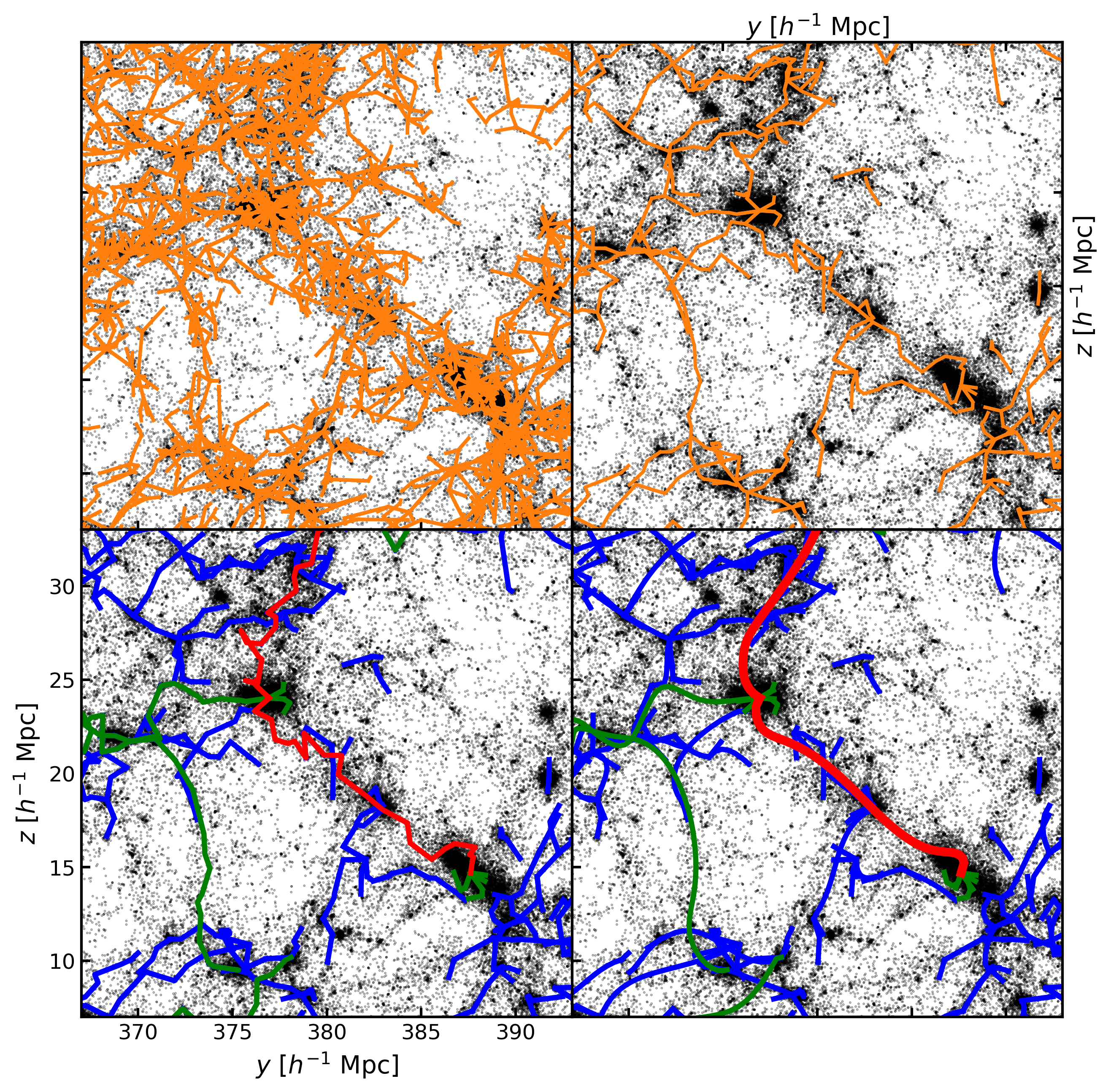}
   \caption{Example of the MST, pruning and smoothing for a $10\ h^{-1}\ \mathrm{Mpc}$ slice from the dark matter only simulation, showing only $1\ \%$ of the particles in this region. Upper-left panel shows the MST corresponding to the slice meanwhile in the upper-right panel shown the pruned MST. 
   Lower-left panel shows the three types of filaments individualised
   (type-2, -1 and -0 in red,
    green and blue line respectively, 
	see text for details). 
	Lower-right panel shows
	the three types of smoothed and
    individualised filaments.}
   \label{fig:semita_en_accion}
\end{figure}

\begin{figure}
	\includegraphics[width=\columnwidth]{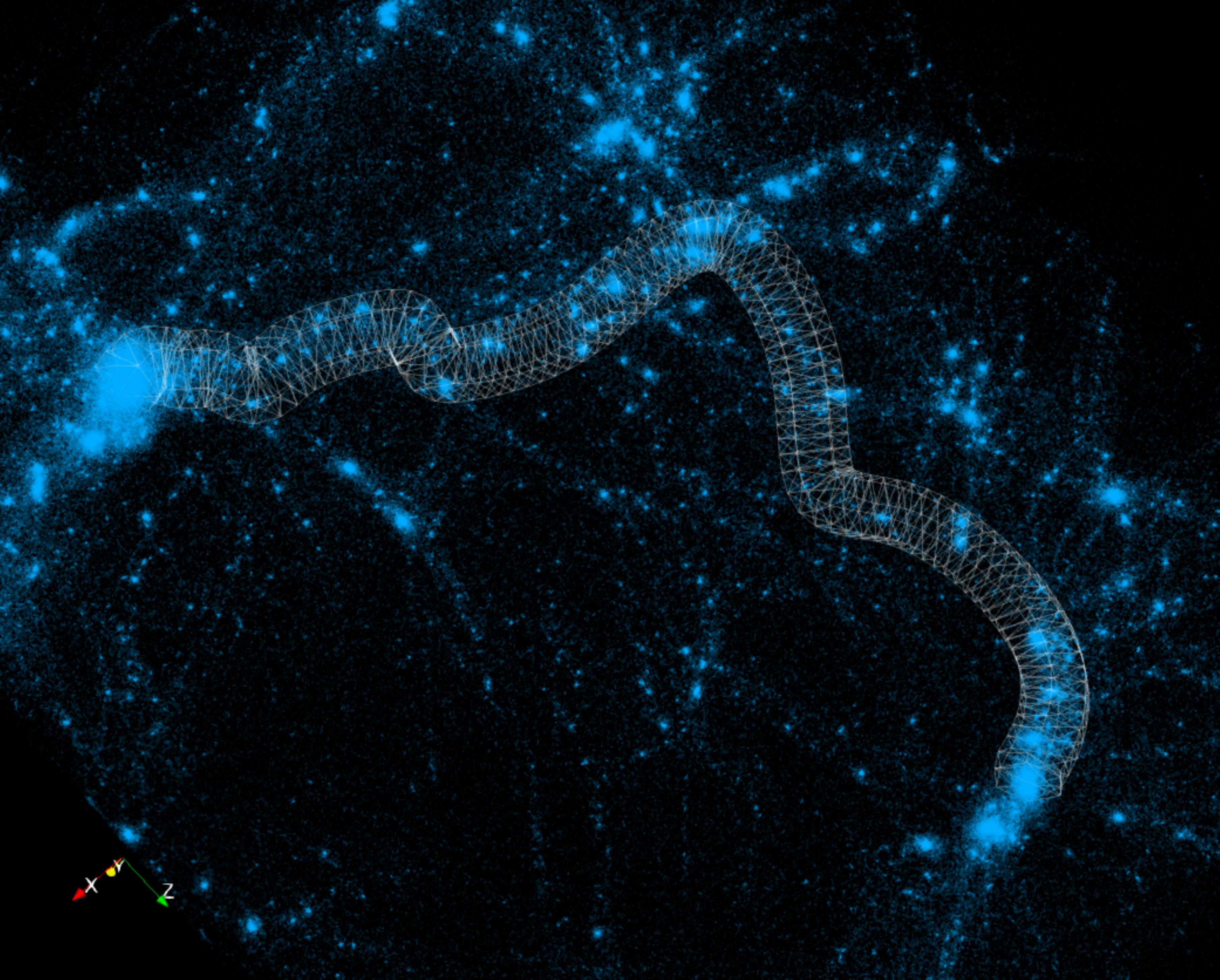}
    \caption{An illustrative example of a filament identified in the last snapshot (z = 0) of the simulation. The total length of this structure is $\approx 30~h^{-1} \mathrm{Mpc}$. The blue dots represent particles of dark matter, while the grey lines show the surface that encloses a tube with a radius of $2~h^{-1} \mathrm{Mpc}$ around the spine of the filament.}
    \label{fig:fila}
\end{figure}

\subsection{Pruning, classification and smoothing of filaments}

Most of the massive halos reside in the knots of the cosmic web and, consequently, are connected by the most prominent filaments. \citep{Colberg2005, AragonCalvo2010b}.
Therefore, we select those halos with masses greater than $M_{th} \, = 10^{14}\, h^{-1} \mathrm{M_{\odot}}$ as the ends of our filaments \citep{Gonzalez2010, Libeskind2018}. 
Given that we are interested in filaments that connect the more prominent halos, we apply a ``pruning'' process that allows us to remove the minor branches that do not belong to the main path that connects these objects.
To do this we follow a similar procedure as described in \citet{Park2009} and \citet{Bonnaire2020} so that we will say that a tree is ``pruned'' to level $p$ when all $k$-branches with $k \leq p$ have been removed. In this work, we pruned our MST to level $p=4$. 
It should be emphasised that we avoid pruning those branches that contain a halo with a mass greater than or equal to $M_{th}$.
Due to this restriction, values larger than $p = 4$ do not cause significant modifications.

Once the pruning process is applied, we individualise the different branches according to the mass of the halos at their ends. To do this, 
we first trim those paths of the MST that connect two halos with masses greater than $M_{th}$ (type-2 filaments). Next, from the remaining tree, we extract those paths that connect the nodes of degree $1$ with the ones whose mass are greater than $M_{th}$ (type-1 filaments). 
Finally, all remaining branches are classified as type-0 filaments, which can be interpreted as ``tendrils'' \citep{Alpaslan2014, Alpaslan2016, Odekon2018}.

At this point we already have a catalogue of individualised filaments, however, since they are forced to pass through all the nodes, 
the filaments do not follow a smooth path. 
Consequently, we use a order $3$ B-spline fitting routine from the FITPACK library \citep{Dierckx1993} to define the spine of the filament.
This is a smooth curve that connects the halos at the ends and passes close to all the intermediate nodes in the filament path.

Summarising, 
in order to run our filament finder it is required to select two percolation lengths for the FoF algorithm, one associated with the identification of tracers ($l_1$) and the other associated with an intermediate density region used for restricting the MST ($l_2$).
The latter parameter allows us to control the typical environment density where the filaments are found. However, it can be changed in a wide range of values depending on the type of filaments of interest.
The third required parameter is the threshold ($M_{th}$) for the mass of the end halos of the filaments, which is used to extract and classify filaments from the pruned MST. In this paper, we use the value
$M_{th} = 10^{14}\, h^{-1} \mathrm{M_{\odot}}$
to restrict our analysis to cluster-cluster filaments. However, another value of this parameter can be chosen, depending on the subject of study, to extract another filamentary pattern such as tendrils.
Another fixed parameters is the pruning level ($p$) of the MST, which is set to $p = 4$ since it is a suitable value to be used in a wide range of applications (we refer the reader to \citet{Bonnaire2020} for a complete study of this parameter behaviour).
Finally, two fixed functions correspond to the weight of the edges, assumed as a Newtonian form, and the method to obtain smooth filament spines in our case a B-spline algorithm.

A schematic view of our identification method can be seen in figure \ref{fig:semita_en_accion}, where we show some steps of the algorithm applied in a $30 \times 30 \times 10 \,\ h^{-1}\ \mathrm{Mpc}$ slice of the simulation described in section \ref{sec:data}. 
For clarity, we only show 1\% of dark matter particles (black points) in this region.
In the upper-left panel, we show the MST (orange lines), can be seen that massive halos have more edges close to them than lower mass halos which is a consequence of our choice for the weighting scheme.
The upper-right panel emphasises the role of pruning the MST.
In general, it can be seen that the main filamentary structure is maintained, while most minor branches have been removed.
Lower-left panel shows the three types of filaments individualised according to the mass of the halos at their ends; red, green and blue lines correspond to the type-2, -1 and -0 filaments, respectively.
Finally, the lower-right panel shows the final result after the smoothing process.

As an illustrative example, figure \ref{fig:fila} shows a typical type-2 filament identified by our algorithm. In this plot, blue points represent dark matter particles while the wireframe represents a tube of $2~ h^{-1} \mathrm{Mpc}$ radius that follows the spine of the filament. As can be seen, this outlines a tangled path through the density field with two massive structures at the ends and smaller halos in its trajectory. This is a typical example that shows how our method links the positions of dark matter halos maximising the matter content inside the tube.

\section{Properties of filaments} \label{sec:properties_filaments}

\subsection{Data}
\label{sec:data}

In this work we use the last snapshot ($z = 0$) of a dark matter
only simulation of $1600^{3}$ particles 
in a periodic box of $400 \, h^{-1} \mathrm{Mpc}$ side 
with cosmological parameters $\Omega_{m} = 0.31$ and $\Omega_{\Lambda} = 0.69$ and $h = 0.68$ given by Planck Collaboration results \citep{Planck2018}.
The particle mass resolution is $1.18 \times 10^{9} \, h^{-1} \mathrm{M_{\odot}}$,
the force resolution is $7.68 \, h^{-1} \mathrm{Kpc}$
and the normalisation parameter is $\sigma_{8} = 0.811$. 
The simulation was evolved using the public version of GADGET-2 code \citep{Springel2005}.

We identified approximately $6.8 \times 10^{6}$ 
dark-matter halos with at least 20 particles
using a FoF algorithm with a linking length $l_{1}$ (see \S \ref{sec:fof}).
The resulting halos distribution is characterised by 
a minimum mass of $ M_{min} = 2.36 \times 10^{10} \, h^{-1} \mathrm{M_{\odot}}$, 
and a maximum mass of $ M_{max} = 1.57 \times 10^{15} \, h^{-1} \mathrm{M_{\odot}}$.
We have a total of 22574 halos with $ M > 10^{14} \, h^{-1} \mathrm{M_{\odot}}$.
The mass contained inside $l_{2}$ regions represents approximately 70 \% of the total mass.

\subsection{Global properties of filaments}

In order to characterise our filaments, we define four basic properties for each filament: mass, length, curliness, and a parameter $q$ that represents the ratio between the masses of the end halos. 
The latter is important since, as shown below, it is an essential parameter to determine the position of the saddle point, which in turn has a key role in various aspects of structure formation such as, for example, the acquisition of the angular momentum of the halos \citep{Codis2012, Codis2015}.

\begin{figure}
        \begin{subfigure}[b]{0.5\linewidth}
        \centering
        \includegraphics[width=\linewidth]{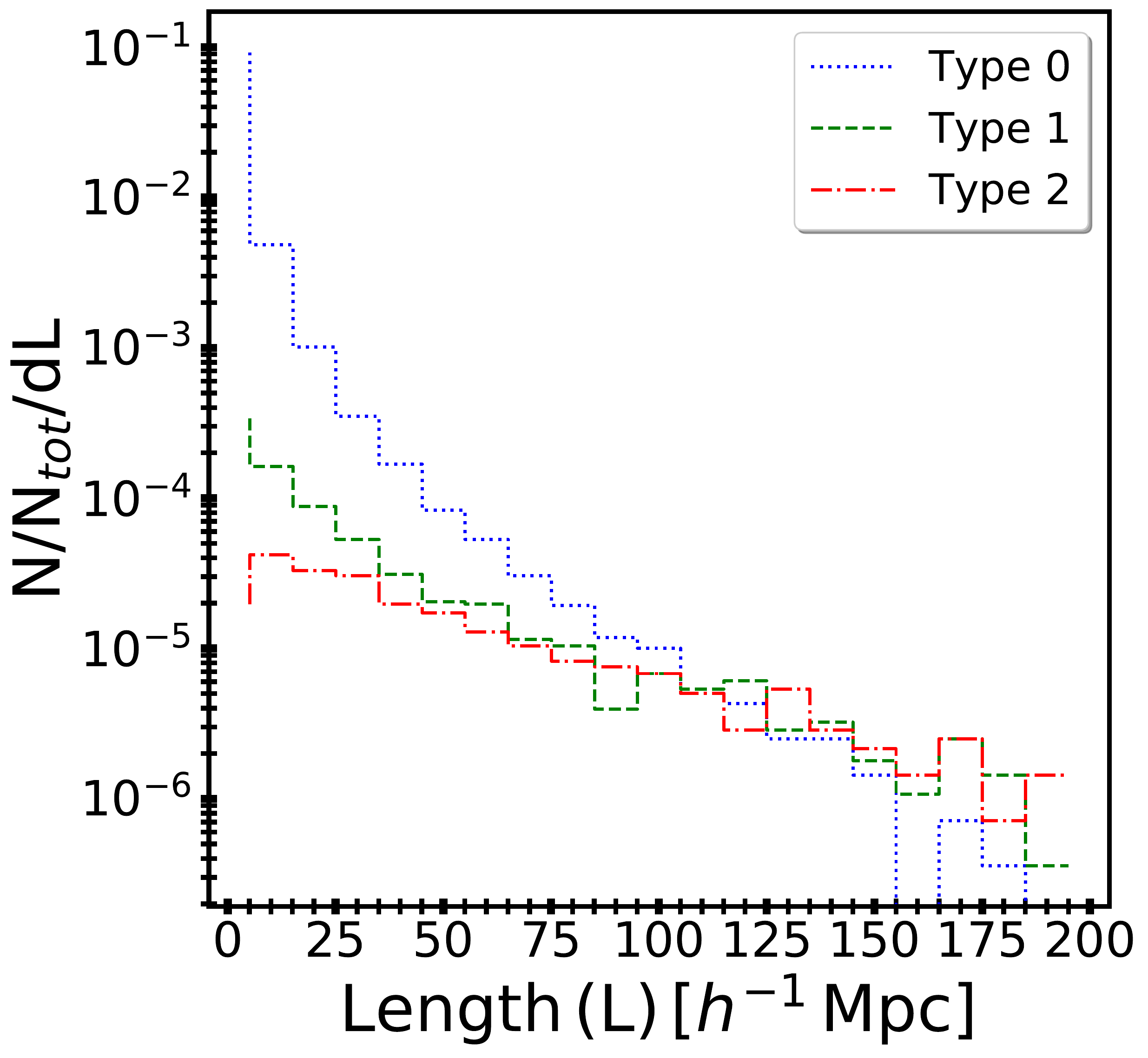}
        \label{fig:Length_Sphere}
        \end{subfigure}%
        \begin{subfigure}[b]{0.5\linewidth}
        \centering
        \includegraphics[width=\linewidth]{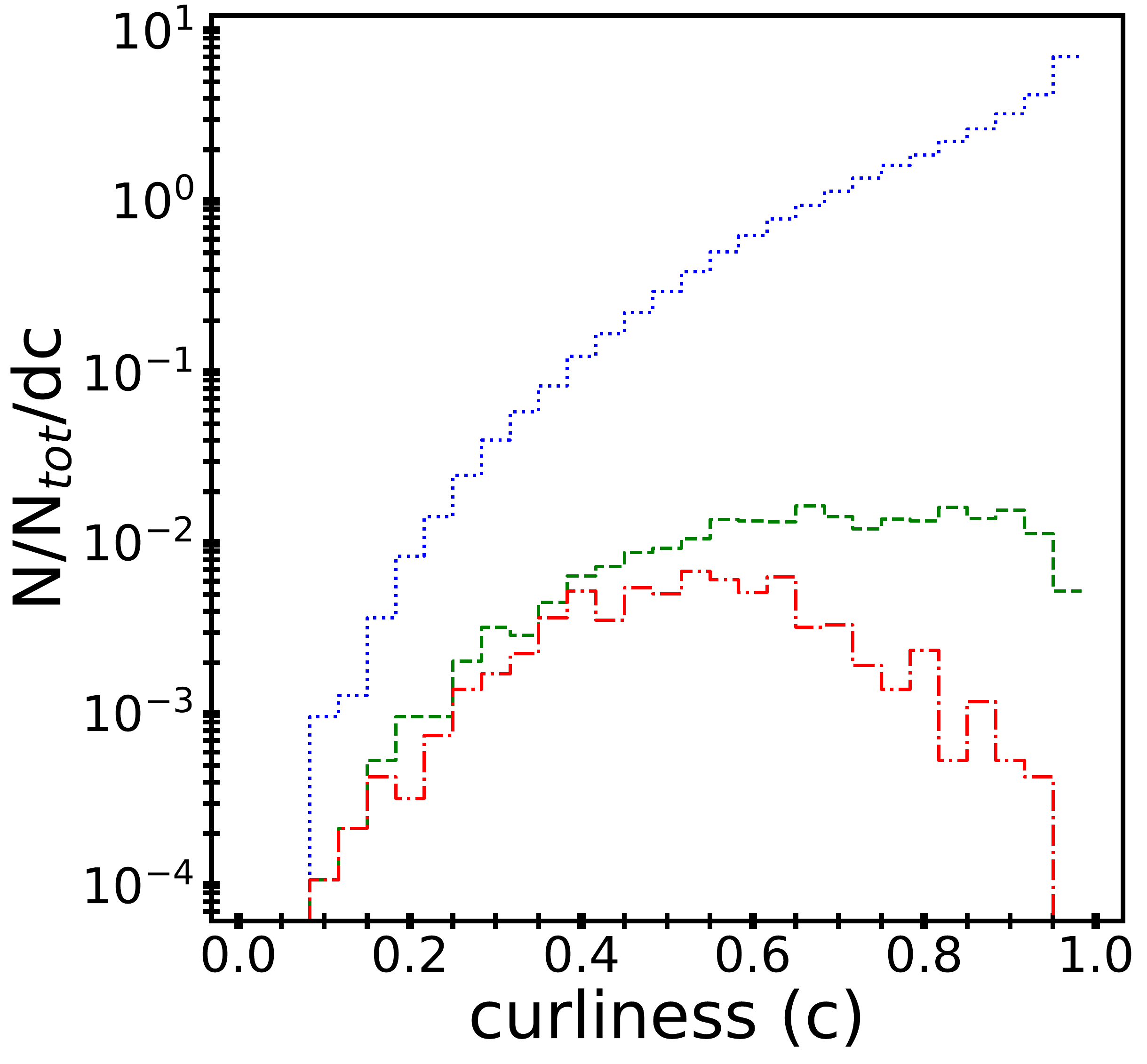}
        \label{fig:Hist_elong}
        \end{subfigure}
        \begin{subfigure}[b]{0.495\linewidth}
        \centering
        \includegraphics[width=\linewidth]{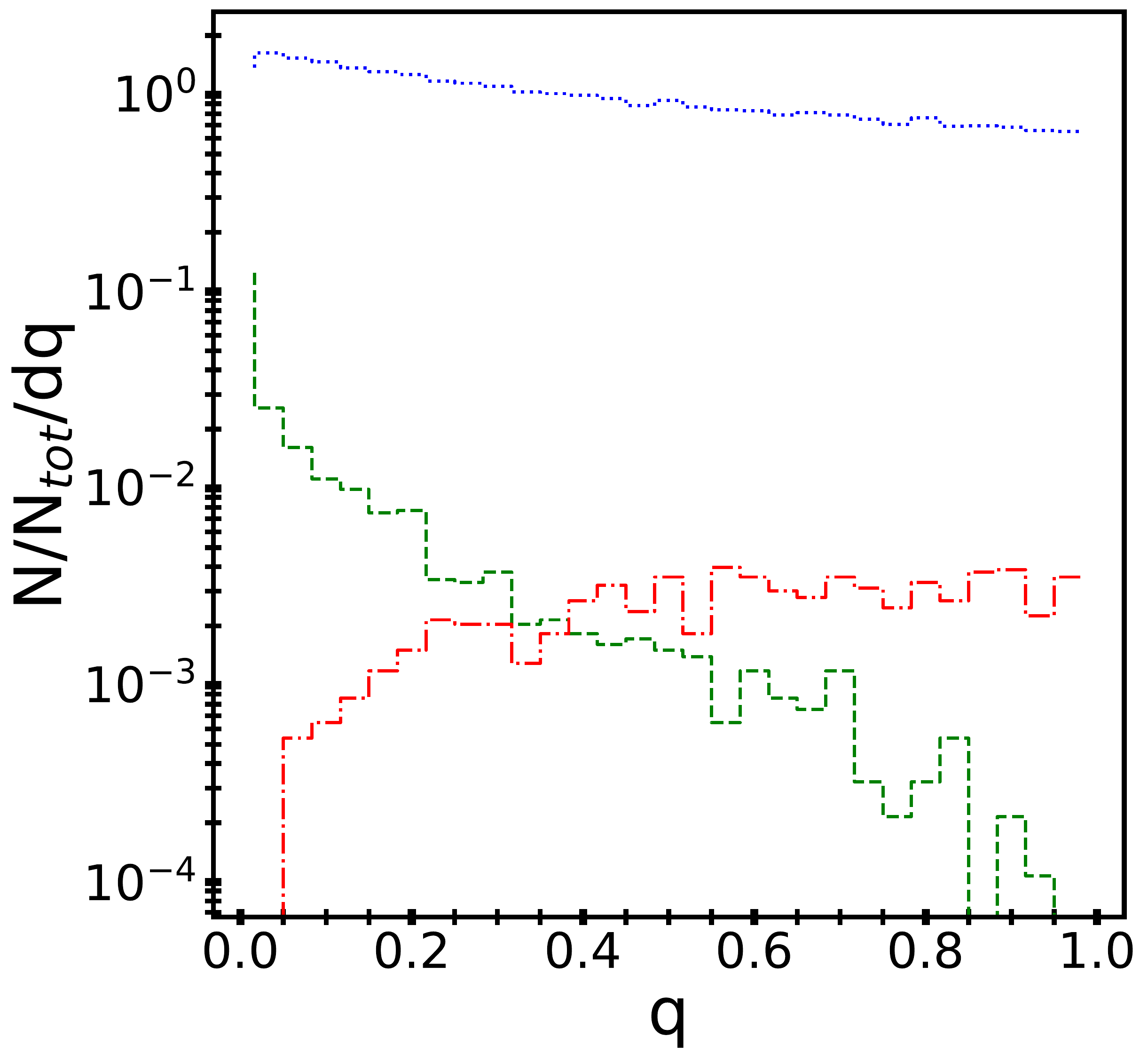}
        \label{fig:Hist_q}
        \end{subfigure}%
        \begin{subfigure}[b]{0.48\linewidth}
        \centering
        \includegraphics[width=\linewidth]{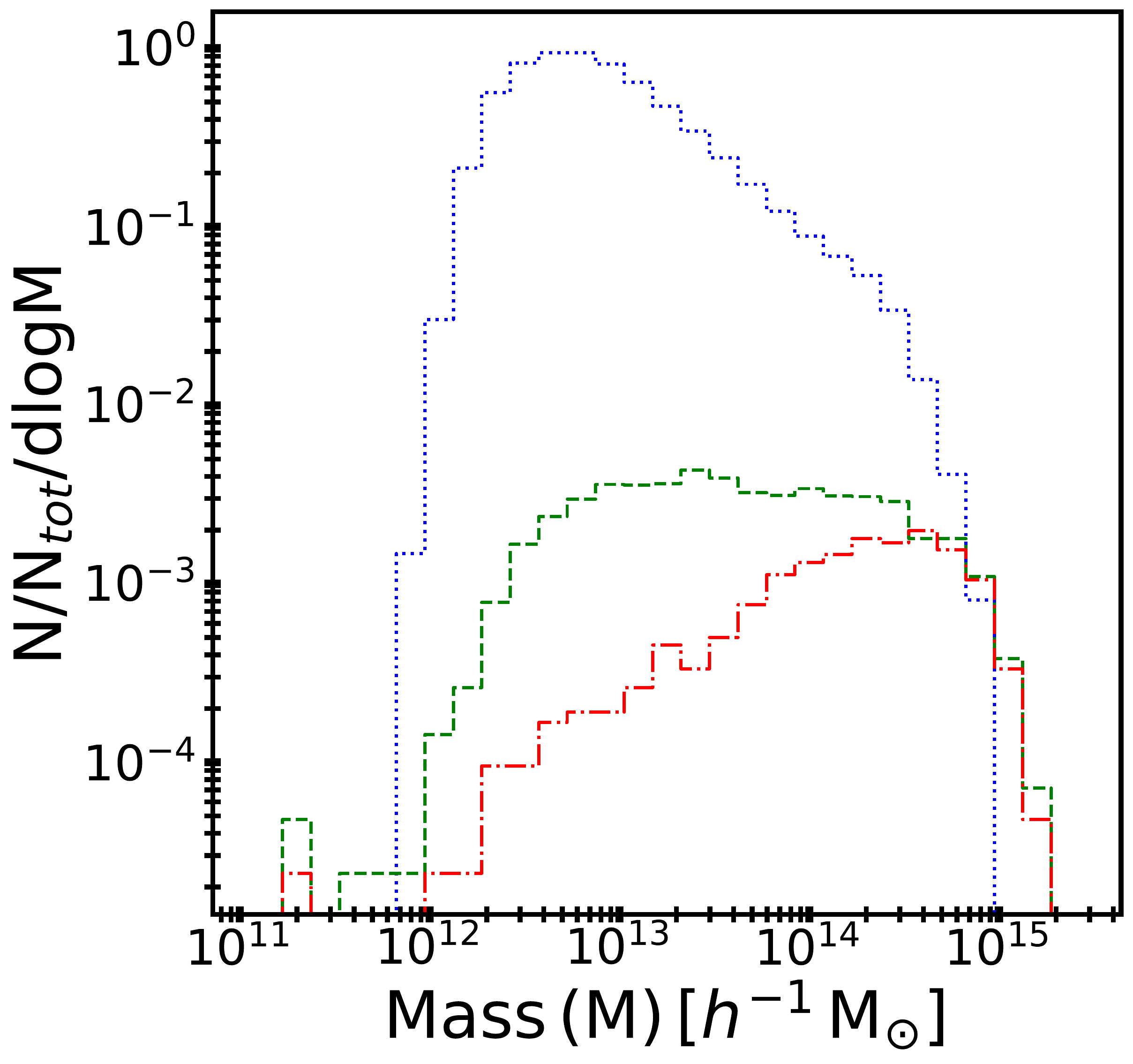}
        \label{fig:Mass_Sphere}
        \end{subfigure}
        \caption{Global properties histograms: 
                 distributions of length, curliness, $q$ parameter,
                 filament mass properties
                 normalized by total filaments number in upper left, upper right, bottom left and bottom right panels for the three types of filaments (type-2, -1 and -0 in red dot-dashed, green dashed and blue dotted line, respectively).
				}
        \label{fig:Histograms}

\end{figure}

The length ($l$) is the most basic and general property of the filaments, and we define it as the sum of the lengths of all smoothed edges that constitute the filament.
The upper-left panel of fig. \ref{fig:Histograms} shows the length distributions for filaments 
type-0, -1 and -2 in blue, green and red colours respectively. 
Given the hierarchical nature of the large scale structure formation, it is expected that type-0 filaments dominate in number as shown in the figure.
The distributions (type-0, -1 and -2) exhibit the same exponential behaviour found by \cite{Bond2010b, Tempel2014a} for the SDSS data set and \cite{Gonzalez2010} for dark matter simulations.

The next property is the curliness ($c$), 
defined as the ratio of the distance between 
the ends and the length of the filament. 
By construction, this is a dimensionless parameter, 
and its values range between $0$ and $1$. 
The straight filaments have an approximate 
curliness value of $1$, 
whereas twisted filaments have values close to $0$.
The upper right panel of fig. \ref{fig:Histograms} shows 
that type-0 filaments are straighter than the other types. 
This is because these filaments are preferably shorter.

The third basic property we define is the ratio $q = M_{1}/M_{2}$,
between the masses of the halos at the ends of the filament, with $M_{2} > M_{1}$.
We shall see in section \ref{sec:sec_saddle} that
the position of the saddle point along the filament
will be approximately related to this ratio. This can be thought in a idealised picture, as the equilibrium of the dominant gravitational pull of both clusters at filament ends.
The $q$ parameter distribution is shown at the bottom-left panel of fig. \ref{fig:Histograms}. 
As expected the number of filaments with small $q$ is larger for type-0.

Finally, the bottom-right panel of fig. \ref{fig:Histograms} shows 
the filament mass distribution, 
which is estimated by counting the dark matter particles
inside a tube with a radius of $2 \, h^{-1} \mathrm{Mpc}$ along the filament spine.
We decided to use this radius because it corresponds to a change of slope in the radial density profile (section \ref{sec:densidad_field}), this is in good agreement with other authors who found that their filaments have typical radii below $2 \, h^{-1} \mathrm{Mpc}$ \citep{Colberg2005, AragonCalvo2010b, Cautun2014, Galarraga2020}.
To calculate the filament mass we exclude the particles at their end that lies within two virial radii.  
As can be seen from fig. \ref{fig:Histograms}, 
type-2 filaments are more massive than type-1.
This is to be expected since type-2 filaments
have two massive halos 
with mass greater than or equal to $M_{th}$ 
in their extremes and therefore, 
those filaments define the stronger bridges of the cosmic web.

Looking at the histograms of fig. \ref{fig:Histograms}, we can see that type-0 filaments are shorter and much less massive than the other types.
Type-0 filaments tend to be shorter in length, while types -1 and -2 are found to have a flatter length distribution.
Therefore, they probably do not fit our definition of filament (cluster-cluster bridges). Hence will not include them in our subsequent analyses.

\begin{figure}
        \centering
        \begin{subfigure}[b]{0.5\linewidth}
        \centering
        \includegraphics[width=\linewidth]{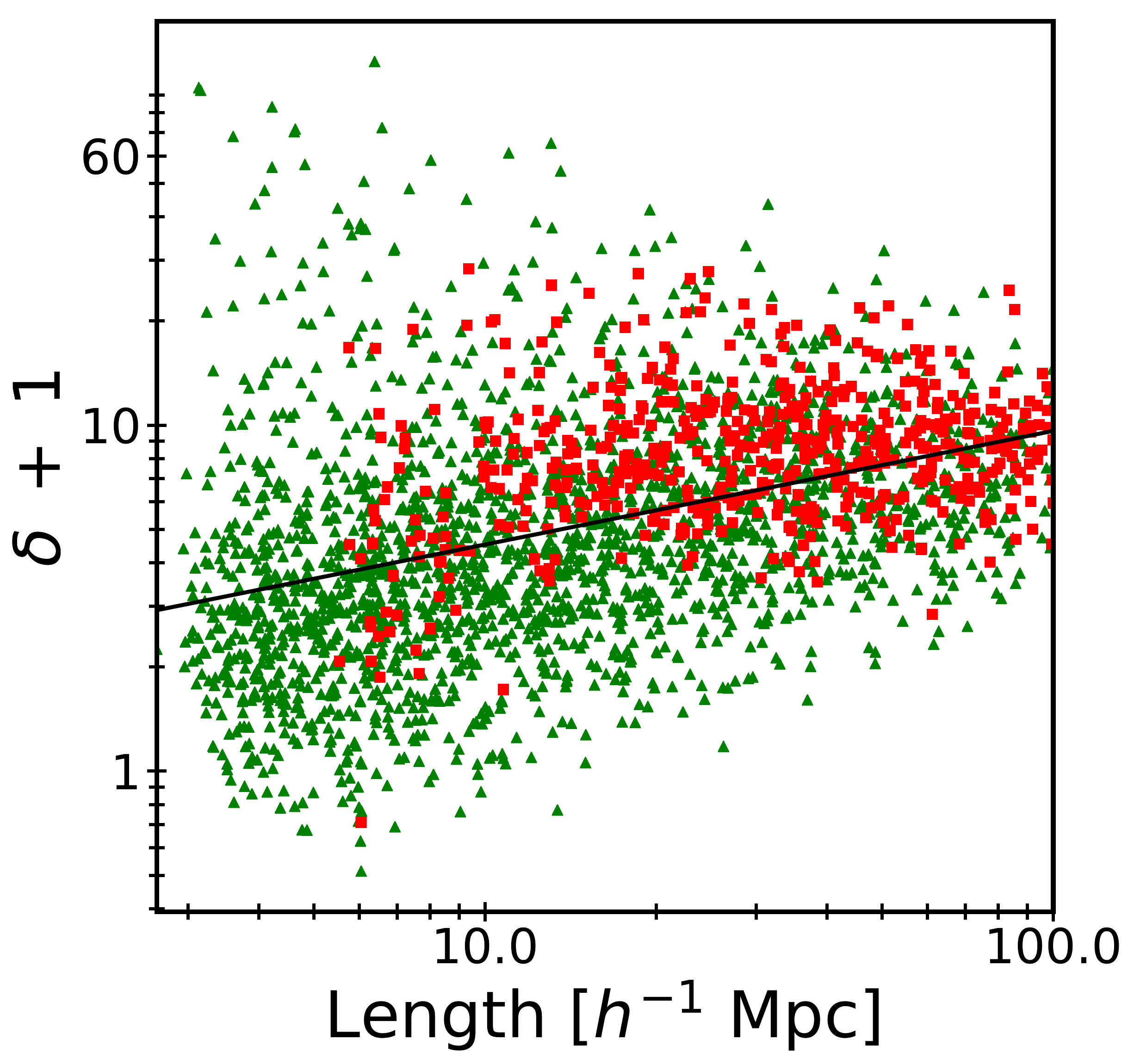}
        \label{fig:Rho_vs_Long}
        \end{subfigure}%
        \begin{subfigure}[b]{0.5\linewidth}
        \centering
        \includegraphics[width=1.025\linewidth]{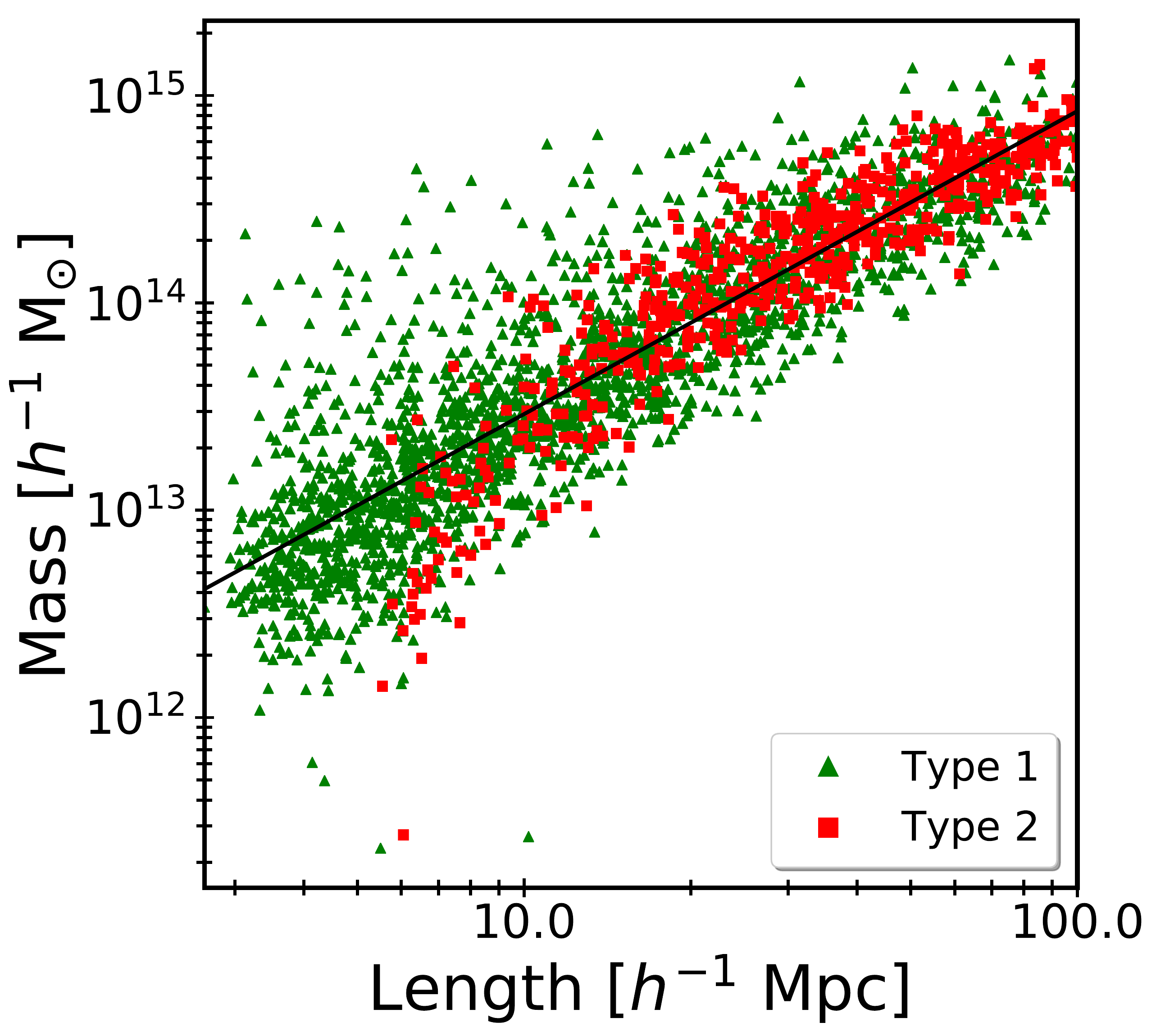}
        \label{fig:Mass_vs_Long}
        \end{subfigure}

        \caption{The relations between length vs. the enclosed overdensity (left panel) and dark matter filament mass (right panel) of type -2 and -1 filaments (red squares and green triangles, respectively) in the sample with length $\in [10, 100] \ h^{-1} \mathrm{Mpc}$. 
        The black solid lines show the best linear fit to the data.}
        \label{fig:Correlation_Properties}        
\end{figure}

An important feature that we can use to assess whether the identified objects are well defined structures is the average overdensity enclosure in a tube with a radius of $2 \, h^{-1} \mathrm{Mpc}$.
In the left panel of fig. \ref{fig:Correlation_Properties}, 
we show the overdensity within the tube as a function of length. The overdensities span a wide range of values ($1 < \delta + 1 < 60$), but are typically around $10$ for type-2 filaments (red dots) and this value is quite independent of the length. This result is similar to that obtained by  \citet{Cautun2014} in the simulations and
\citet{Kuutma2020} in observational data.
On the other hand, type-1 filaments (green dots) present smaller values of $\delta$ since they are typically less massive and they would be less prominent.
For completeness and comparison, we also calculate the relationship between mass and length (right panel of fig. \ref{fig:Correlation_Properties}) which, as expected, shows a linear positive dependence with length having a slope of approximately $1.46$. Although this slope does not match that of \citet{Cautun2014} ($M \propto L^{2.2}$), it should be taken into account that the way of calculating the masses are quite different. 
These authors estimate the masses by adding the contributions of the ``voxels'' that are part of their filaments according to the criteria of their identifier, while we do it by adding the contained mass within a fixed distance ($2 \, h^{-1} \mathrm{Mpc}$) to the axis of the filament.
Furthermore should be taken into account that our sample could be biased respect to that of \citet{Cautun2014} given that we are only considering type -1 and -2 filaments, which, as we will see later, are comparable to the thickest ones of their sample.

\section{Dynamical and structural properties of filaments}
\label{sec:dynamical_filaments}

In this section, we study the average density 
and velocity fields of the filaments by means of a stacking technique.
The most natural way to do this is to fix the ends of the filaments and normalise their lengths. 
More precisely, the coordinates of a particle tracer around a given filament will be given by two distances adapted to the geometry of the spine of the filament.
One coordinate, denoted with $r$, is the distance from the tracer particle to its closest point in the spine.
The other coordinate $z$, is the distance from this point to the less massive halo measured along the filament.
Therefore, the stacking technique simply consists in normalising $z$ to the length of the filament $l$ and averaging all the physical quantities in bins in both $z/l$ and $r$ coordinates.
Since we are interested on volume-weighted quantities, we need to compute the volume of each bin. However, due to the wavy nature of the filaments, each bin do not have a regular shape in the comoving coordinates of the simulation, consequently, the estimation of its volume can not be done analytically. Therefore, we employ a Monte Carlo technique, using a random sample $100$ times denser than the mean density of the simulation.

As previously mentioned, type-2 filaments are the most significant bridges in the large scale structure, therefore, for the sake of clarity, hereafter we concentrate our studies on these filaments, even through, the results for type-1 filaments are similar.
We will focus on filaments with length between $10$ and $100 \, h^{-1} \mathrm{Mpc}$. 
The upper limit is given by the size of the simulation and the lower limit set to avoid the effect produced by the massive halos at the filament end.

\subsection{Density field}
\label{sec:densidad_field}

\begin{figure}
	\includegraphics[width=\columnwidth]{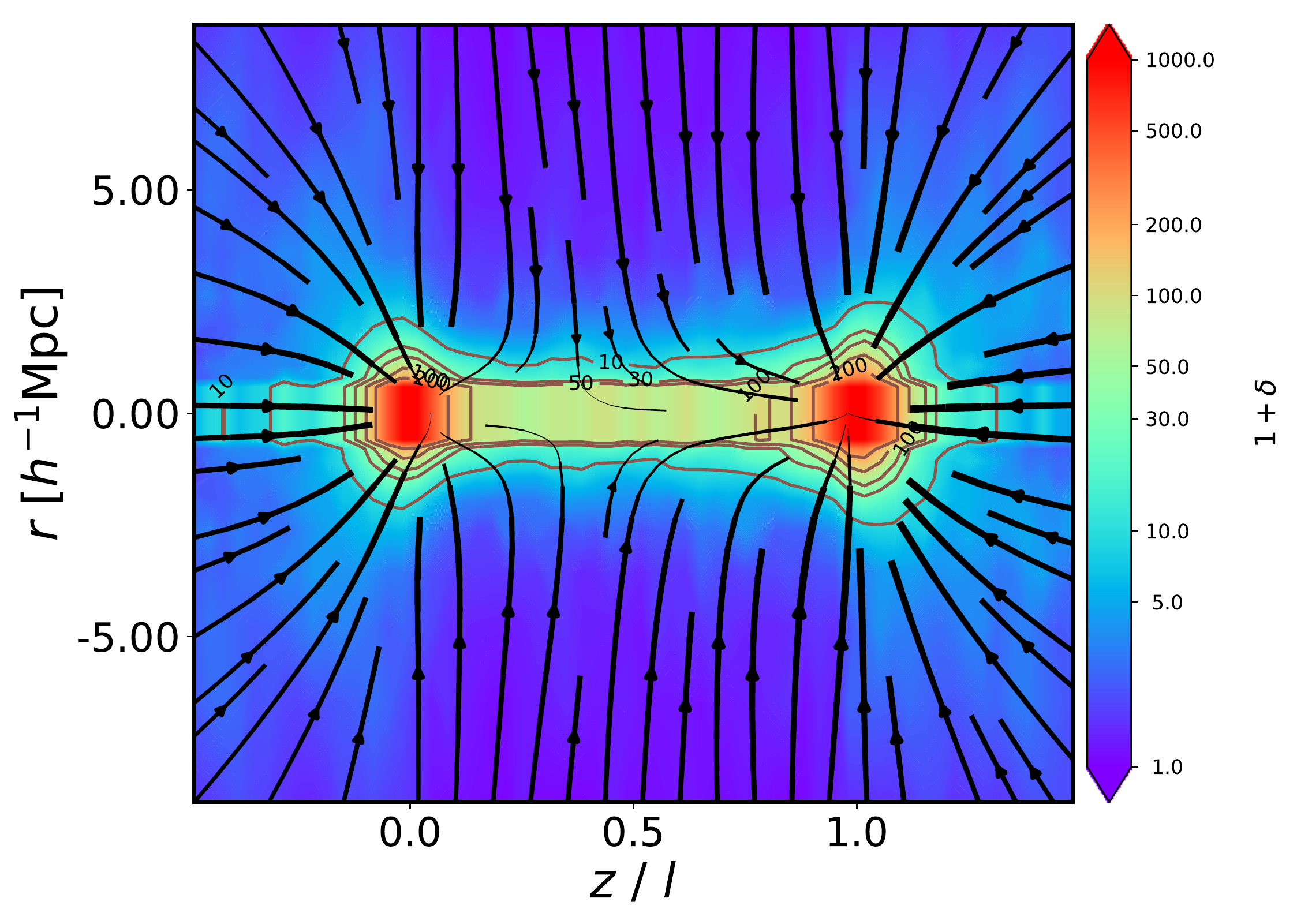}
	\includegraphics[width=\columnwidth]{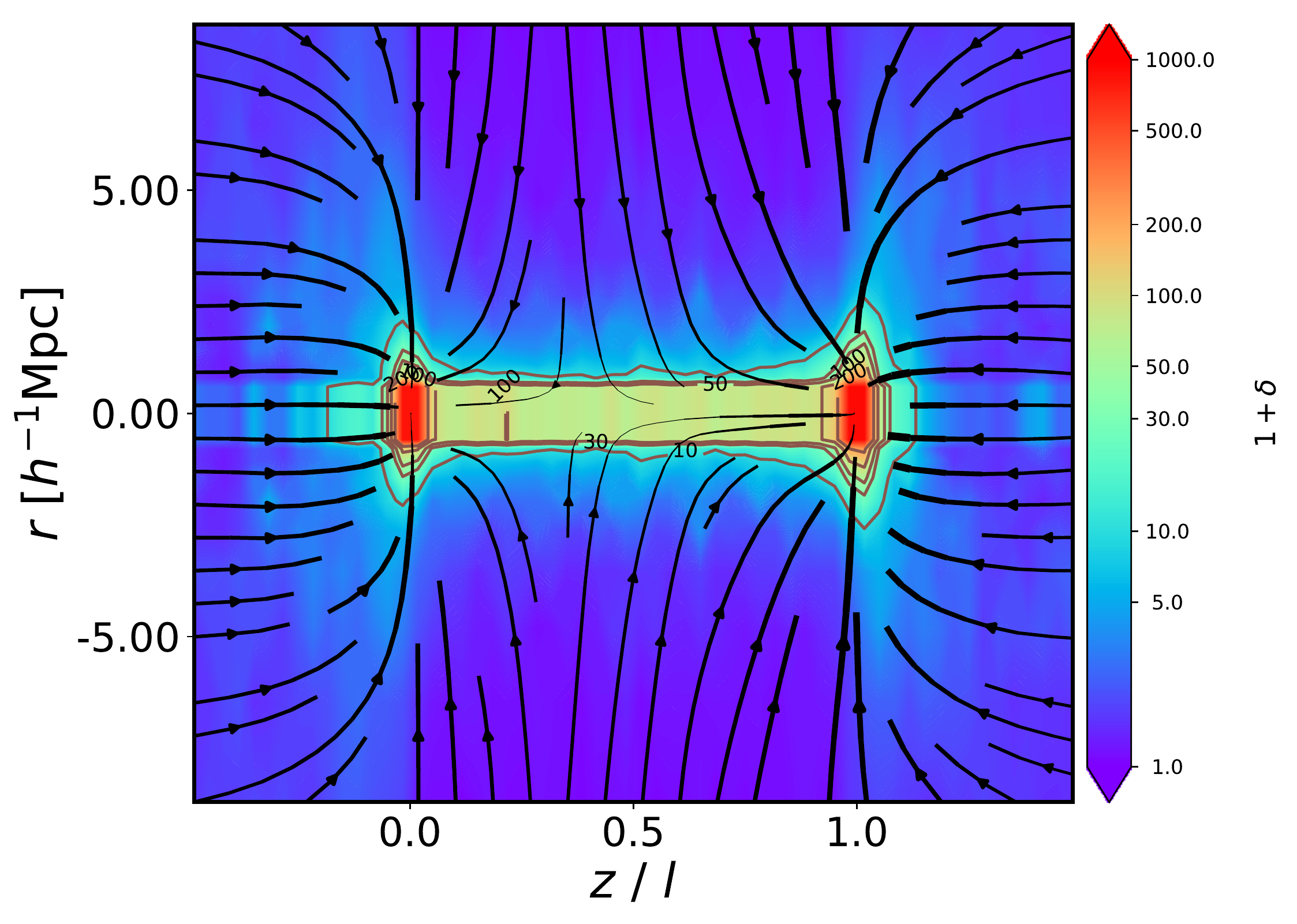}
    \caption{
			Normalise density stacking for filaments with 
			length $\in [10.00, 19.21] \ h^{-1} \mathrm{Mpc}$
			and $(30.88, 50.64] \ h^{-1} \mathrm{Mpc}$ 
			in top and bottom panel, respectively.
			The streamlines represent the velocity field.
		}
    \label{fig:huesos}

\end{figure}

The normalised stacking plot for the overdensity 
shown in fig. \ref{fig:huesos} represents
the typical shape of a filament \citep[e.g.][]{Kraljic2019} 
with high-density peaks at the extremes, 
indicating the position of clusters, 
and a matter bridge linking them. 
The coloured contours represent the overdensity levels
and the streamlines show the velocity field
with its thick proportionally to the velocity magnitude. 
We left the analysis of the velocity field 
for the next subsection, here,
we will concentrate on the density field.
It should be mentioned that, for aesthetics, 
the stacking plot is reflected along the $z$ axis.

The two density peaks at the extremes of the filament 
are the most noticeable features in this figure.
The nature of the stacking method leads the filaments to always have a massive halo at the ends, therefore it is expected for the signal to increase at these points.
Meanwhile, the bridge of matter between them is formed by halos less massive than $M_{th}$ and diffuse matter whose signal is averaged in the stacking process.
Even though the filaments are non-virialised and irregular structures, 
it can be observed that an overdensity of $\delta + 1 \approx 10$ 
encloses almost the entire filament structure, 
in good agreement with \cite{Cautun2013}.
In the direction perpendicular to the filament, the overdensity reaches values as high as $\delta + 1 = 100$ near the spine, decreasing to values of
approximately $\delta + 1 \approx 1$ at distances greater than $5\ h^{-1}\ \mathrm{Mpc}$.

The signal shown on the left and right outside the filament reflects the nature of the cosmic web, in the sense that there are no isolated filaments, and the extremes are connected with other filamentary structures, which are diluted when stacked.

\begin{figure}
	\includegraphics[width=\columnwidth]{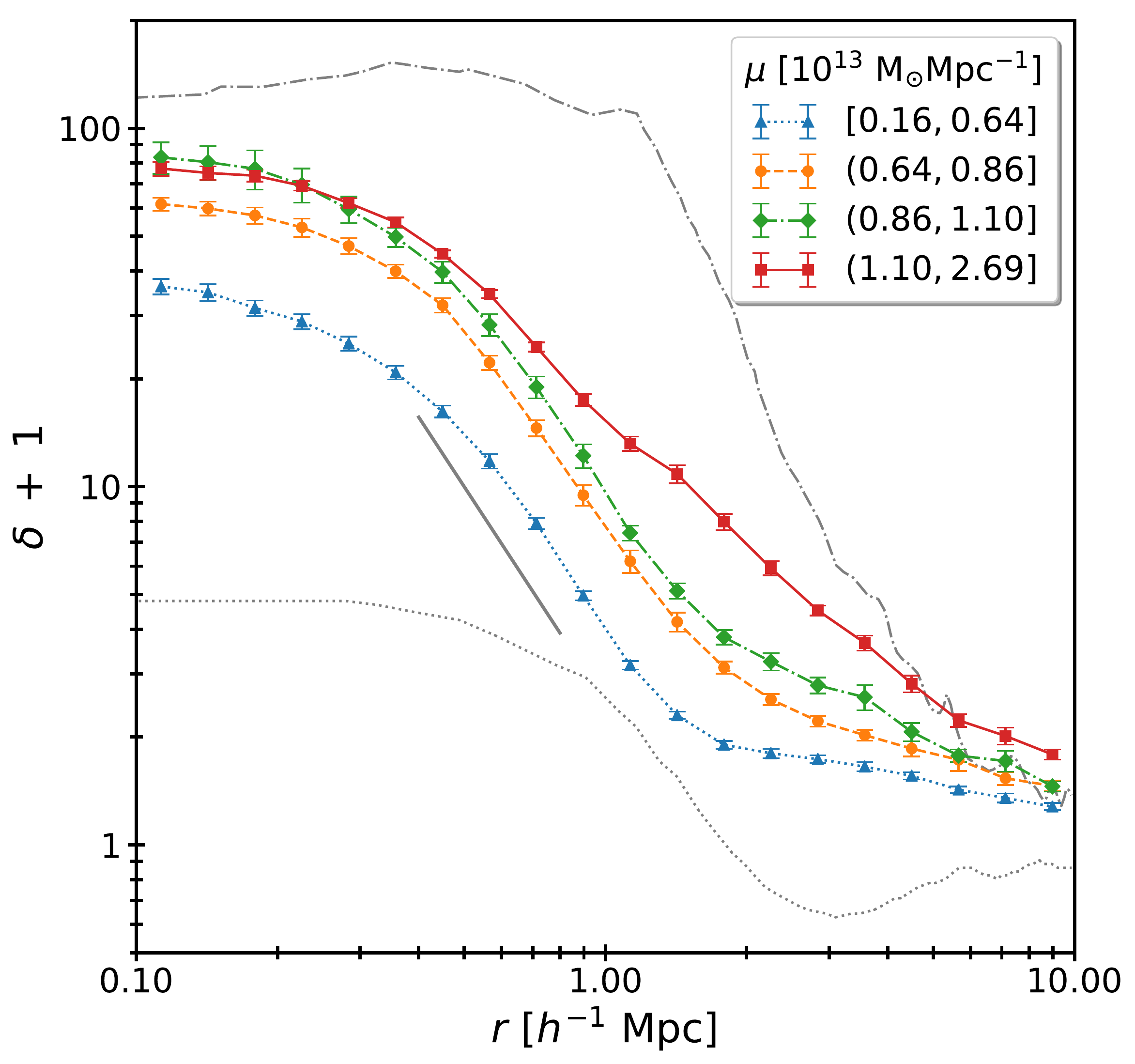}
    \caption{
		The radial density profiles of cosmic filaments: 
		Overdensity  $\delta + 1$
		as a function of the distance $r$ from the spine of filament
		in quartiles of the linear density ($\mu = M/l$).
		The profiles of \citet{Cautun2014} are shown in dot-dashed and dotted grey curves  
		for the $(6 - 8)$ and $(1 - 2) h^{-1} \mathrm{Mpc}$ diameter filament samples, respectively.
		The grey solid line correspond to the $r^{-2}$ power-law. 
	}
    \label{fig:profile}
\end{figure}

In addition to the normalised overdensity stacking we also calculate the radial density profile.
This is estimated in concentric cylindrical shells around the spine of the filament in the coordinate space $z/l-r$. 
Figure \ref{fig:profile} shows the density profile at $20$ equal logarithmically spaced bins between $0.1$ and $10\, h^{-1} \mathrm{Mpc}$. 
The different curves represent the profile estimated for $4$ subsamples of $\mu$, where $\mu = M/l$ is the mass per unit length. It is worth remembering that in the estimation of the filament mass the halos at the ends are excluded.
The error bars are computed using the jackknife technique.
The filaments with larger $\mu$ are denser towards the centre, while they are also wider.
In agreement with previous works \citep{Colberg2005, Dolag2006, Gonzalez2010, AragonCalvo2010b} 
in intermediate scales ($0.5$ to $2.0 \, h^{-1} \mathrm{Mpc}$), the profile approximately follows a $r^{-2}$ power-law (solid grey line).
As expected, on large scales, all the curves tends to the local background density of the Universe independently of $\mu$. 
It should be noted that despite the difference of amplitude between the four profiles, all shapes are similar.
Comparing our density profile with those obtained by \citet{Cautun2013} for the filament samples of larger ($6 - 8\ h^{-1} \mathrm{Mpc}$) and smaller ($1 - 2\ h^{-1} \mathrm{Mpc}$) diameter in dot-dashed and dotted grey lines respectively, it can be seen that our filaments show similar behaviour to their larger diameter samples. Nevertheless, our profiles are sharper towards the centre.
As stated at the beginning of this section, we are showing results only for type-2 filaments, however it should be stressed that similar profiles to those showed in fig. \ref{fig:profile} are obtained for type-1 filaments with slightly smaller amplitudes.
Our profiles also show a flat behavior near the center. This behavior may not correspond to a physical property of the filaments, but rather could be related to the smoothing technique in our method or to the difficulty of establishing a well defined axis in this type of irregular objects.
Likewise, comparing the profiles with recent results in the large hydro-dynamical simulations
\citep{Galarraga2020}, we find that they are qualitatively similar. It is worth noting that these profiles reach higher values of overdensity towards the filament axis. This can be attributed to the identification through the DisPerSE algorithm, whose definition very well traces the filament spine since it follows the gradient of the density field.

\begin{figure}
		\includegraphics[width=.97\linewidth]{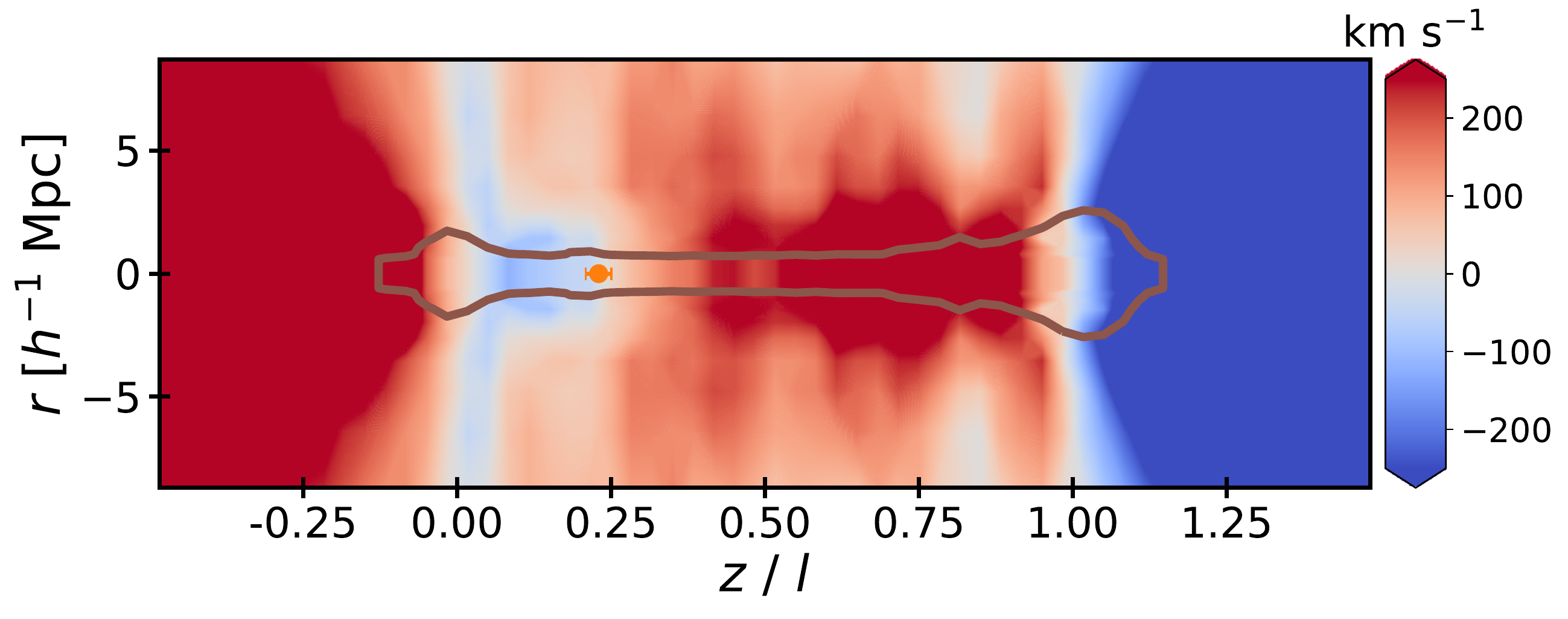}
        \includegraphics[width=.97\linewidth]{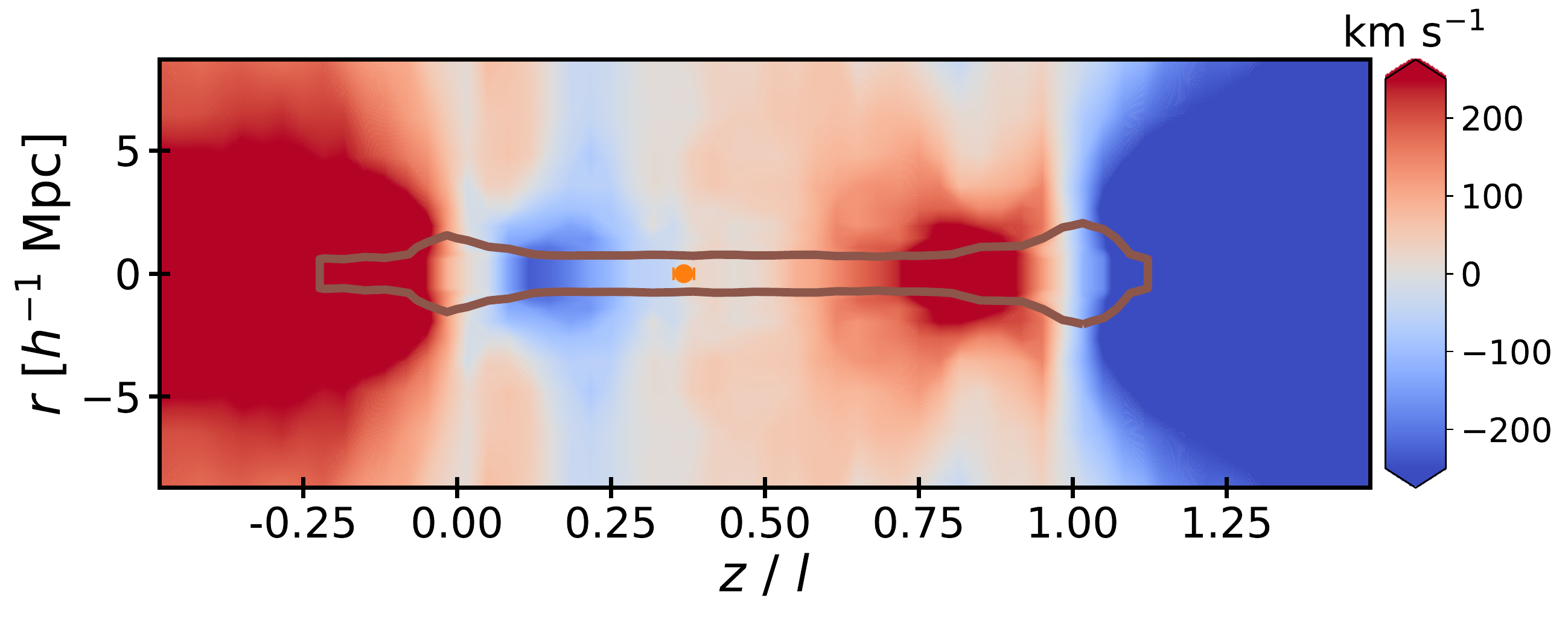}
        \includegraphics[width=.97\linewidth]{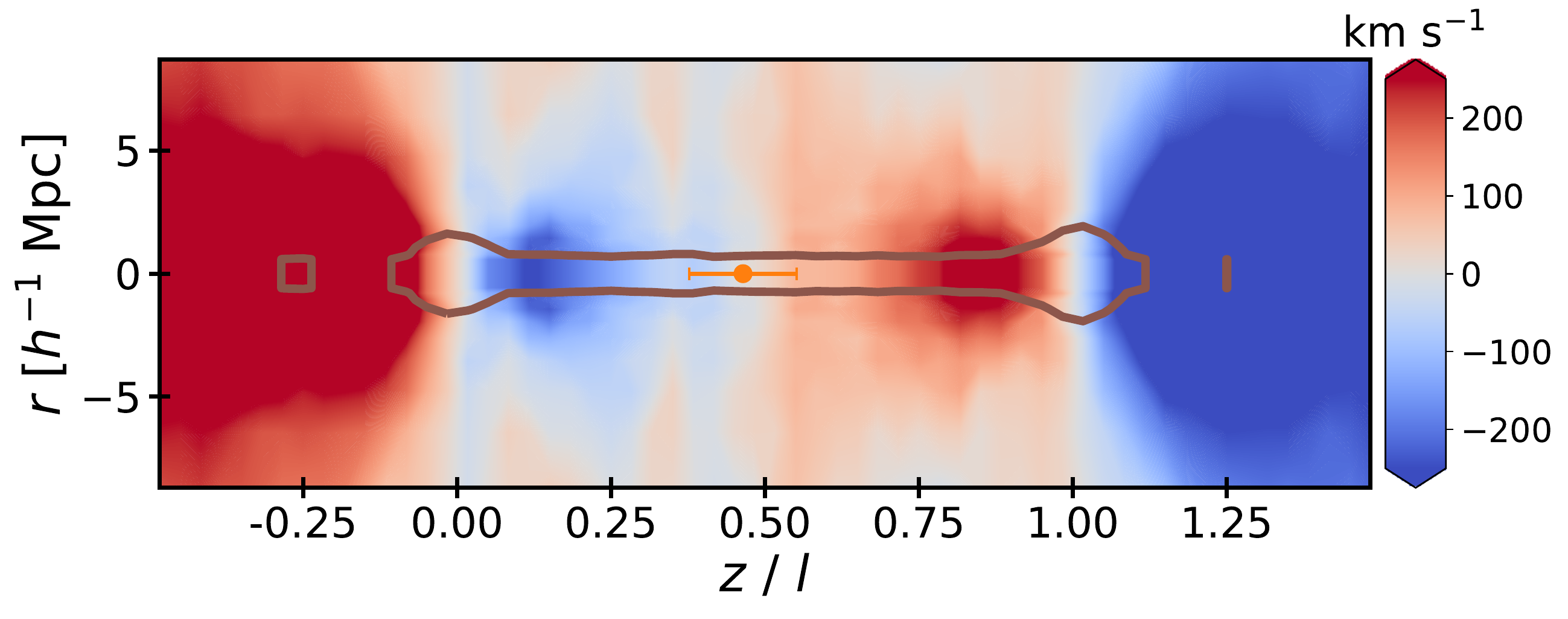}
        \includegraphics[width=.97\linewidth]{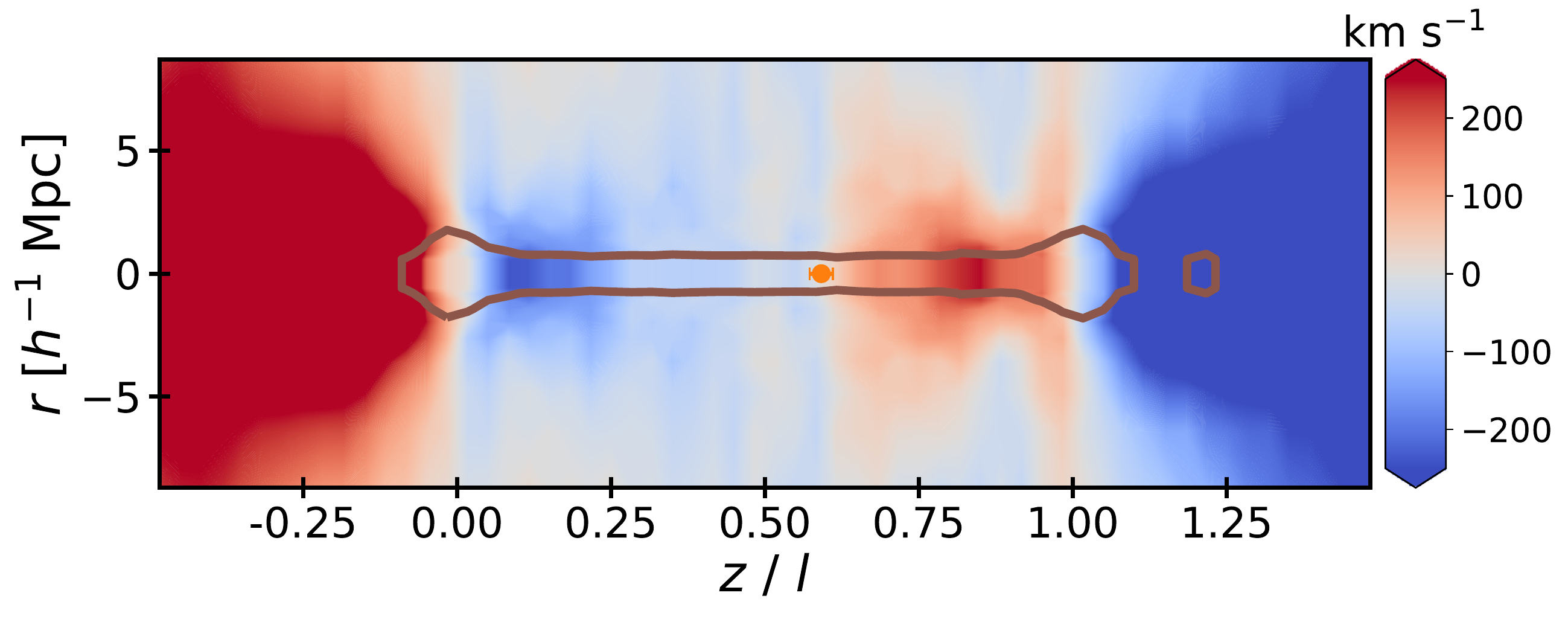}
        \caption{Parallel component of the mean velocity field 
		in quartiles of the $q$ parameter
		for filaments with length 
		$\in (19.21, 30.88] \ h^{-1} \mathrm{Mpc}$.
		The blue and red colours
		represent the negative and positive
		velocity component, respectively. 
		The brown contour represents the iso-overdensity $\delta + 1 = 10$. The orange dot represent the position of the saddle point.}
    \label{fig:vel_parallel_by_q}

\end{figure}

\begin{figure}
    	\includegraphics[width=.97\linewidth]{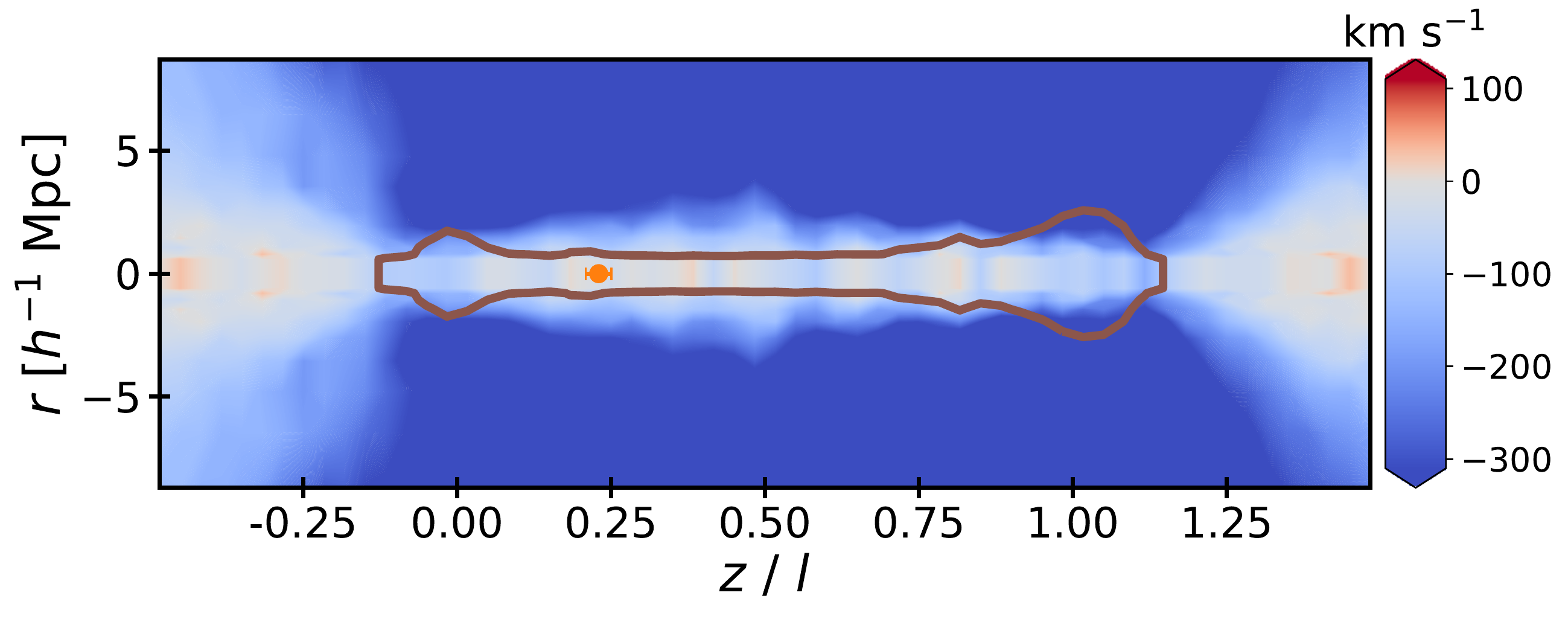}
        \includegraphics[width=.97\linewidth]{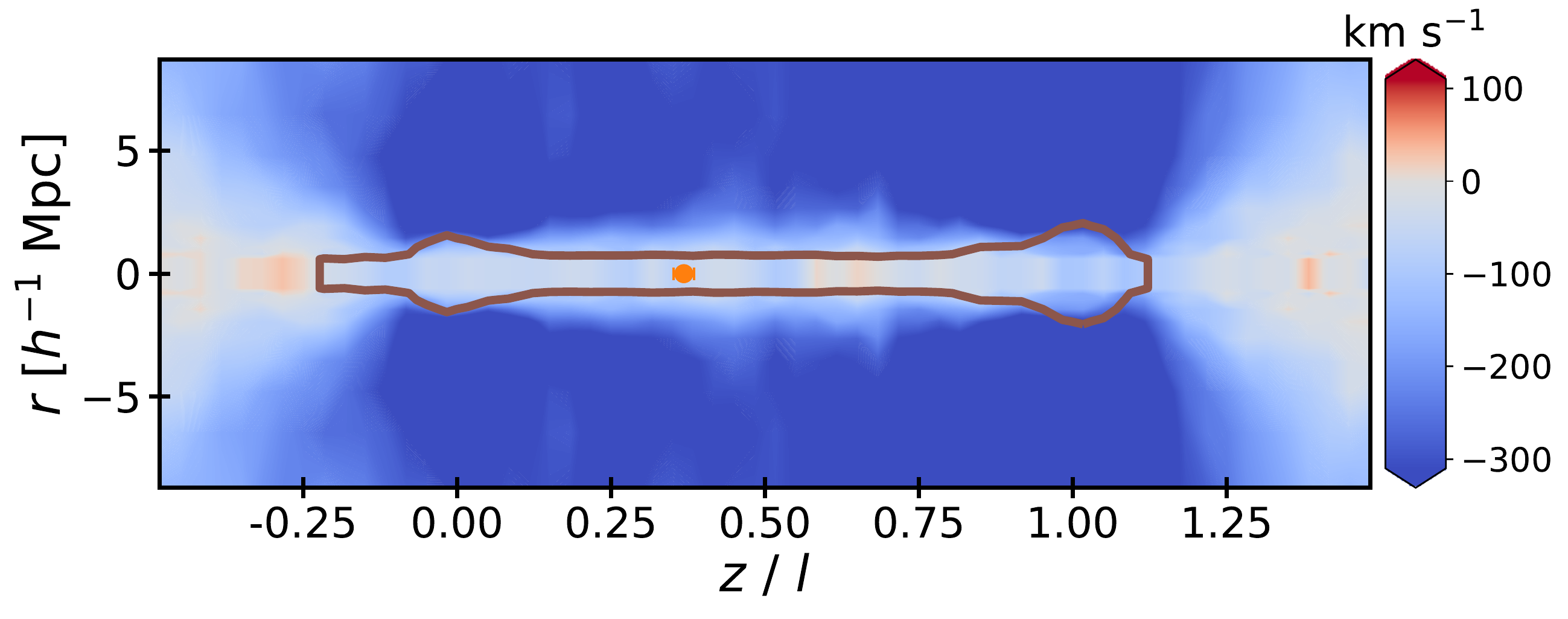}
        \includegraphics[width=.97\linewidth]{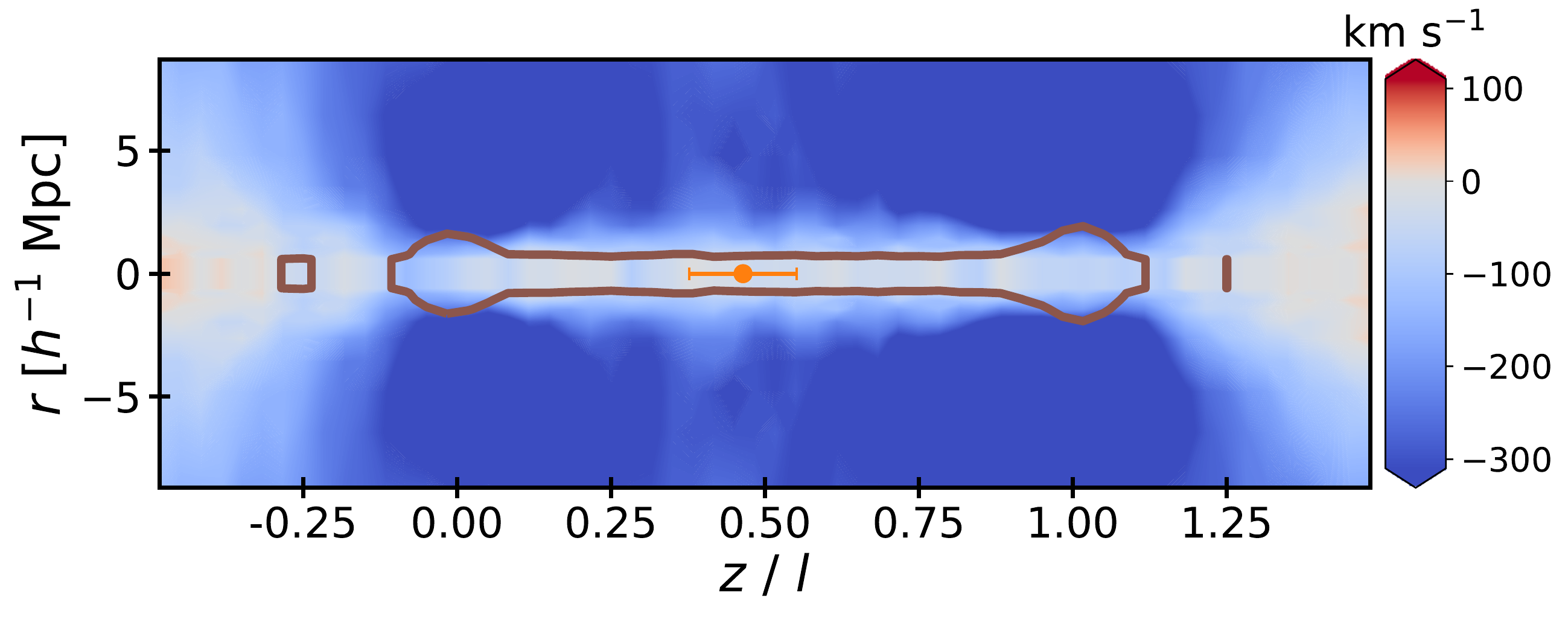}
        \includegraphics[width=.97\linewidth]{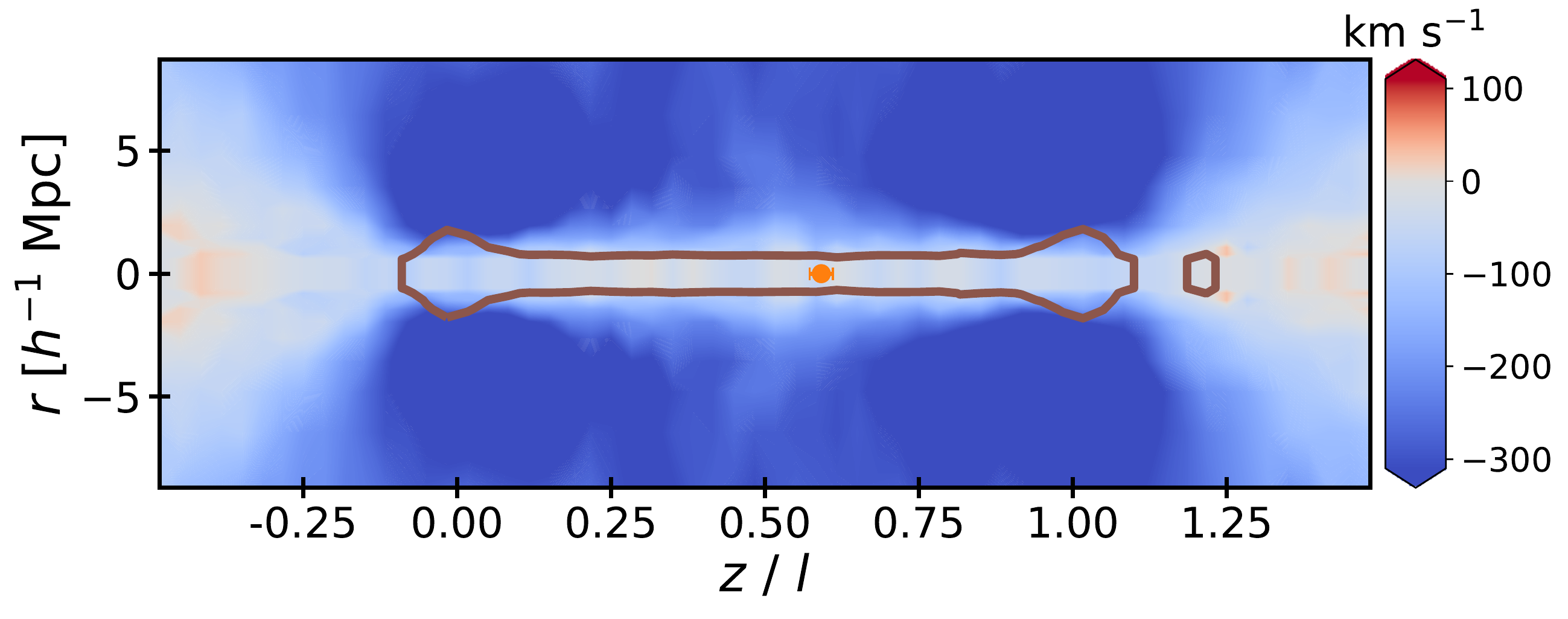}        
        \caption{Perpendicular component of the mean velocity field 
	    		in quartiles of the $q$ parameter
		    	for filaments with length 
    			$\in (19.21, 30.88] \ h^{-1} \mathrm{Mpc}$.
	    		The blue and red colours
		    	represent the negative and positive
			    velocity component, respectively. 
			    The brown contour represents the iso-overdensity $\delta + 1 = 10$. The orange dot represent the position of the saddle point.}
    \label{fig:vel_perp_by_q}

\end{figure}

\subsection{Velocity field}

The dynamical processes involved in filament formation have a key role in the genesis of properties of dark matter halos and galaxies \citep[e.g.][]{Jones2010, Codis2015, Laigle2018, Kraljic2018, Kraljic2019}. This motivates us to characterise the velocity field around the filaments and, to do so, we will separate it into its parallel, $v_{\parallel}$, and perpendicular, $v_{\perp}$, components respect to the spine of the filament. In the case of the perpendicular component, is define positive in the direction away from the filament axis, while the parallel component is positive in the direction that points to the most massive halo.
In order to subtract bulk motion, we refer all velocities to that of the centre of mass of the halos at the extremes of each filament.

As mentioned previously, the streamlines in fig. \ref{fig:huesos} show the stacked velocity field. There it can be seen that at large distances matter falls perpendicularly to the filaments and as it gets closer, its velocity decreases in magnitude and becomes parallel. In agreement with the scenario presented in previous works \citep[e.g.][]{Bond1996, Colberg2005, AragonCalvo2010b, Kraljic2018, Kraljic2019}, the flow of particles within the filaments moves towards the ends and the saddle point (where the streamlines diverge towards the extremes) can be neatly distinguished. 

Figure \ref{fig:vel_parallel_by_q} shows the parallel component of the average velocity field in quartiles of the $q$ parameter. The blue and red colours represent negative and positive velocity components, respectively. For the sake of clarity, the brown line shows the contour level of overdensity $\delta + 1 = 10$. 
The orange point indicates the position of the saddle point. The error bars are calculated using the jackknife method. We determine the location of the saddle point on the axis of the filaments, as the place where the velocity flux diverges towards the maximums, that is, where a sign inversion of the velocity vector occurs along the parallel direction.
As expected, the magnitude of the velocities increase as the material moves towards the nodes and, as can be seen, the saddle point shift from the left towards the centre as the value of the parameter $q$ increases, i.e., as the mass of the halos at the ends resemble each other.

If the filaments are over-density regions where the gravitational collapse occurred towards one-dimensional structures, then it is expected that they generate a potential with the same geometric characteristics and, therefore, 
we should expect that, in the linear regime, the perpendicular component of the velocity increases towards the axis of the filament.
The averaged perpendicular component of the velocity field is shown in 
fig. \ref{fig:vel_perp_by_q} where it can be seen that although at great distances there is a fall towards the filament when we approach to the axis, 
the average perpendicular velocity decreases due to the beginning of an anisotropic collapsing process. 
Another characteristic that can be observed in this figure is that the perpendicular infall does not depend on the parameter $q$, which would indicate that the gravitational potential would be determined by the mass of the filament itself and not by the halos at the ends. The high infall speeds away from the filaments are consistent with the expansion velocities at the boundaries of empty voids \citep{Ceccarelli2013}.

\subsubsection{Transverse velocity dispersion}

\begin{figure}
	\includegraphics[width=\columnwidth]{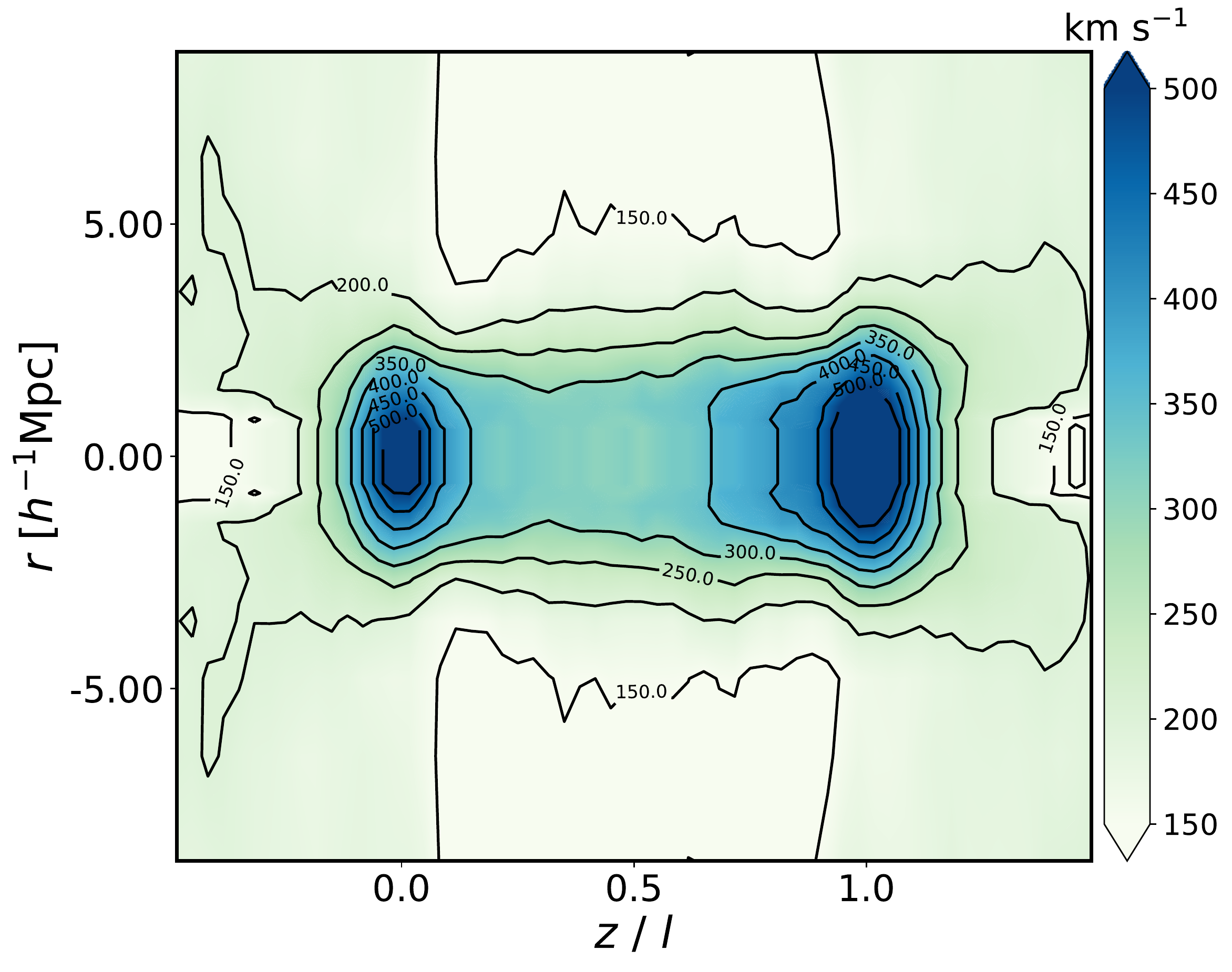}
    \caption{
    Normalise transverse dispersion velocities stacking for filaments  
    with length $\in [10.00, 19.21] \ h^{-1} \mathrm{Mpc}$.
    }
    \label{fig:Lenght_10_07_19_21_LOG_std_per}
\end{figure}

As stated above, the filaments are structures that are in the process of formation and, therefore, have not yet reached dynamic equilibrium. In the sense that they are not supported against collapse by velocity dispersion as such is the case of dark matter halos. 
However, towards their centre, it can be seen that the \textit{velocity dispersion perpendicular} to the axis is greater than in the surrounding regions, as can be seen in fig. \ref{fig:Lenght_10_07_19_21_LOG_std_per}.
This effect can be explained by considering two processes:
(i) Filaments have a certain degree of virialisation in the direction perpendicular to them. That is, anisotropic collapse predicts a well-defined evolution, with regions first contracting into flattened walls, following by elongated filaments and only then collapsing in each direction. In all cases, the structures are supported by velocity dispersion only in already collapsed directions \citep{Hahn2015, Buehlmann2019}.
(ii) The velocity dispersion is due to the encounter between the material accreted into the filament and material making its first shell-crossing.
In this complex scenario, both processes could be happening simultaneously. It should be noted that the filaments contain diffuse material as well as virialised halos.
The velocity dispersion values ($\sigma \approx 100 - 500\ \mathrm{km}\ \mathrm{s}^{-1}$) obtained are in agreement with  \citet{Buehlmann2019}, who independently study the velocity dispersion field.

In order to understand the nature and dynamic of filaments,
\citet{Eisenstein1997} propose a theoretical relation to estimate, from observational data, the mass per unit length $\mu$ of filaments. Specifically, assuming that filaments are axisymmetric, isothermal structures virialised along the perpendicular direction, they found an analytical relation between $\mu$ and the transverse velocity dispersion $\sigma_{\perp}$ of the filaments.

\begin{equation}
    \mu = \frac{{\sigma_{\perp}}^{2}}{G} = 3.72 \times \ {10}^{13} \mathrm{\ M_{\odot} \ {Mpc}^{-1}} \ {\left(\frac{\sigma_{\perp}}{400 \ \mathrm{{km\ s}^{-1}}}\right)}^{2}
		\label{eq:eq_mu_sigma}
\end{equation}

\begin{figure}
	\includegraphics[width=\columnwidth]{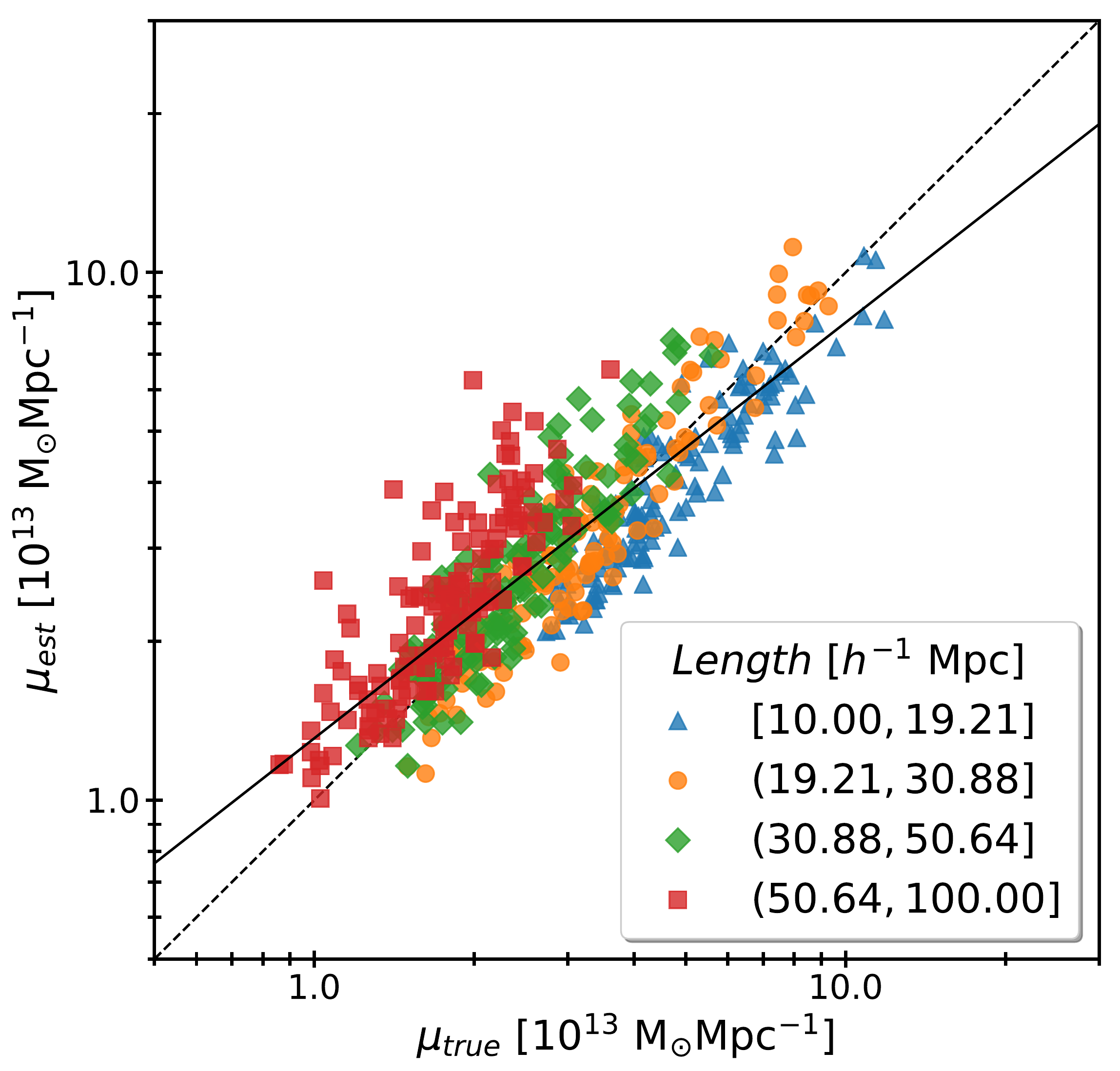}
    \caption{			
			The estimated linear density from the transverse velocity dispersion against the actual linear density for four-length ranges (blue triangles, orange circles, green diamonds, and red squares, respectively). 
			The best linear fit and the identity relation are shown in solid and dashed lines, respectively. 
			}
    \label{fig:mu_est_vs_true_lenght}
\end{figure}

Equation \ref{eq:eq_mu_sigma} differs from equation 13 of \citet{Eisenstein1997} by a factor of $2$, because their equation considers that the velocity dispersion is measured along a line-of-sight.
Figure \ref{fig:mu_est_vs_true_lenght} shows the comparison 
between the true mass per unit length $\mu_{true}$ and the dynamical mass $\mu_{est}$ estimated using equation \ref{eq:eq_mu_sigma}.
The best linear fit and the identity relation are shown in solid and dashed lines, respectively. 
The value of $\mu_{true}$ is calculated as the ratio between the total mass inside a cylinder of radius $2 \, h^{-1}
\mathrm{Mpc}$ along the filament spine and its total length. 
The different colour symbols represent these values for each
type-2 filament on quartiles of length. As it can be seen, the relation is independent of the length of the filament. 

Despite the assumptions made to deduce the eq. 		\ref{eq:eq_mu_sigma}, our results show a good agreement with the theoretical predictions. This would suggest that the filaments are partially virialised in the direction perpendicular to the axis. Strengthening this hypothesis, it is worth mentioning that the crossing time in the perpendicular direction of the filaments is shorter than the Hubble time.
However, the data has a high dispersion, which may be due to the fact that the filaments, in general, do not have a regular cylindrical shape, as assumed to deduce the eq. \ref{eq:eq_mu_sigma}.
In addition, they are not relaxed structures but are continually disturbed by the infall of matter.
Another effect that should be considered is the presence of substructure that tends to overestimate the mass of the filaments \citep{Eisenstein1997}.

\subsubsection{Saddle points}
\label{sec:sec_saddle}

\begin{figure}
	\includegraphics[width=\columnwidth]{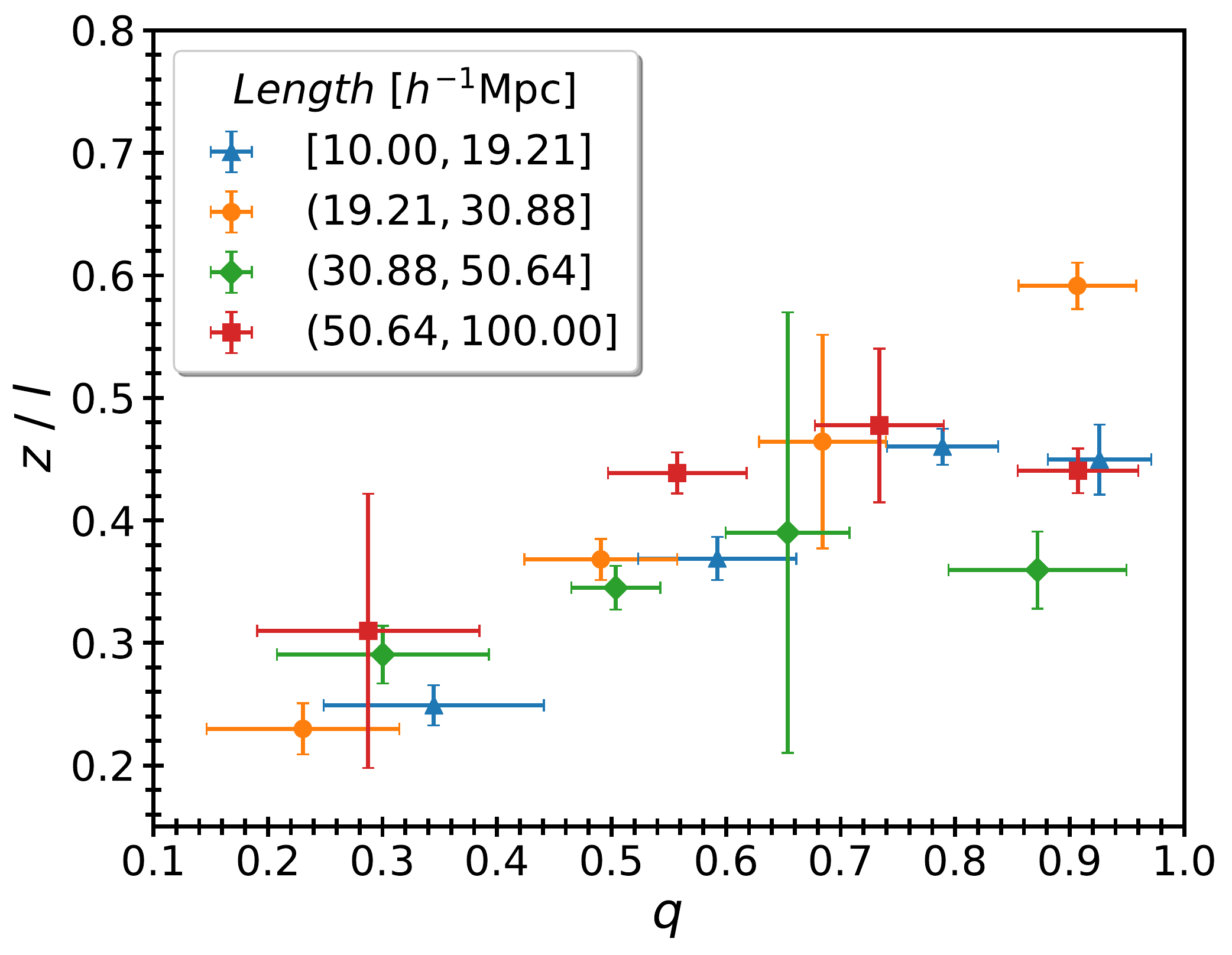}
    \caption
    {
		Correlation between parameter \textit{q} and saddle point position in quartiles of length samples.
	}
    \label{fig:saddle}
\end{figure}

According to our definition of filament, these are a matter bridge that joins two high density peaks. From this definition it follows that at some point in the middle of the peaks there must be a saddle point where the density reaches a minimum and the velocities become divergent. This point plays an important role in the internal structure of the filaments, since it corresponds to the position of dynamic equilibrium along its spine \citep{Pogosyan2009, Codis2015, Laigle2015, Kraljic2019} and that is why a variety of finding algorithms use this feature to identify filaments \citep{Novikov2006, Sousbie2008}. Our method makes no assumptions about this, and yet the saddle point arises naturally as observed in the
fig. \ref{fig:vel_parallel_by_q} and \ref{fig:vel_perp_by_q}.
It is expected that the position of this point depends on the parameter \textit{q} since the dynamics within the filament is affected by the halos at the ends. Figure \ref{fig:saddle} shows the relation between parameter \textit{q} and the distance from the less massive end of the filament to the saddle point for four samples of filaments with different lengths. As expected, there is a direct relationship between them and the saddle point approaches the centre of the filament as the mass of the ends resemble each other. In addition, this behaviour seems to be independent of length.
The bars represent the errors estimated by the jackknife technique and the mean standard deviation for the position and the value of \textit{q}, respectively. The position of the saddle point is estimated by inspecting the averaged parallel velocity field and looking for the point where there is a change in the sign of such velocity component.

\section{Summary and conclusion}
\label{sec:summary_discussion}

In this work, we analyse the properties of the filamentary pattern of a cosmological simulation of $400\ h^{-1}\mathrm{Mpc}$ side and a mass resolution of  $1.18 \times 10^{9} \, h^{-1} \mathrm{M_{\odot}}$, focusing our study on cluster-cluster filaments identified by a novel algorithm developed for this purpose.
Briefly, our finding process consists of successively extracting branches of a Minimal Spanning Tree (MST) constructed from the Friends of Friends halos (FoF) identified in the simulation. 
To avoid false identifications, we restrict the tree to the regions enclosed by an overdensity $\delta = 1$. In this way, each filament is a contiguous path of halos embedded in those intermediate overdensity regions.
To construct the filament catalogue, we additionally prune the MST keeping only those branches with order $k >= 4$ and apply a smoothing filter to the resulting paths.

Our algorithm individualises three types of filaments according to the mass at the ends. If both halos have masses greater than a given mass threshold ($M_{th}$), the filament is classified as type-2. 
We define a filament as type-1 if only one halo at the ends has a mass greater than the threshold, whereas if neither of the two halos exceed $M_{th}$ we denote it as type-0.
Accordingly to previous works \citep{Colberg2005,Gonzalez2010,AragonCalvo2010b}, we use in this work an $M_{th} \, = 10^{14}\, h^{-1} \mathrm{M_{\odot}}$.  

Our results show a good agreement with the general picture of the cosmic web where cluster-cluster filaments are bridges of matter connecting maxima of the density field and delineate the underdense basins.
The stacked density field of these structures shows a ``thighbone-like'' shape surrounded by a typical overdensity $\delta \simeq 10$.

By construction, type-0 filaments are typically short, straight, and more numerous structures of the filamentary network. On the other hand, type-2 filaments are the most dominant bridges of the cosmic web.

We find that the average overdensity within a $2\ h^{-1} \mathrm{Mpc}$ tube along the axis of these filaments has an approximately constant value over a wide range of lengths.
Despite, the dispersion around the mass-length relation is quite large, which may be since filaments do not follow a regular geometric shape, we found that this can be well described by a power-law $M \propto L^{1.46}$. The slope greater than 1 suggests that long filaments are well-defined structures instead of concatenated short filaments.

We measure the transverse density profile of cluster-cluster filaments up to $10 h^{-1} \mathrm{Mpc}$. Finding that filaments with larger linear density are thicker and denser towards the spine of filament. In intermediate scales, the density profile follows a $r^{-2}$ power-law. Meanwhile, on large scales they asymptotically tend to the mean density value. 

In spite of the fact that our method does not impose any restrictions on the velocity field, we are able to determine the location of the saddle point along the axis of the filaments, defined as the place where the velocity flow diverges. The vicinity of the saddle point is a region of particular interest because it is closely related to the rotation of dark matter halos, as well as to the processes of galaxy formation \citet{Codis2015, Laigle2015, Kraljic2019, Lopez2019}. This topic will be investigated in a forthcoming work.

The streamlines pattern of the velocity field shows the ideal scenario where matter flow from low density environments to the filaments, to ultimately infall into the halos at the ends. 
Even though, it should be stressed that our results also indicate that the magnitude of the velocity field along filaments is significantly smaller than velocity field in low density regions. 
In other words, most of the matter is accreted along filaments, however particles infalling into the halos directly form low density regions have the largest velocities \citep{Ceccarelli2011}.
The analysis of the perpendicular component of the velocity field strengthens the idea that the filament produces a cylindrical-like potential since the average velocity values decrease towards the axis while the dispersion increases.

Although filaments are structures in a quasi-linear formation regime, we observed that at their centre the transverse velocity dispersion is higher than in their surroundings. 
This could be indicating that filaments have a certain degree of virialisation. 
In this sense, we analyse the relationship between the transverse velocity dispersion and the linear density. 
We found a good agreement with the theoretical relation proposed by \cite{Eisenstein1997}. 
The observed dispersion can be ascribed to different factors, e.g., filaments are not isolated nor are they smoothed structures, they have a wide range and varieties of substructures.

As a final remark, we stress that, given the flexibility and the few free parameters of our algorithm ($M_{th}$, $l_1$, $l_2$, $p$), the methodology presented in this work is well suited for large cosmological N-body simulations (since we use dark matter halos as structure traces) and for the new generation of spectroscopic surveys. In particular, we believe that the different stacking techniques developed in this work would be useful to overcome the difficulties to obtain the filament properties from galaxy surveys (due to the systematic effects, such as the "Finger of God" effect, edge effects and low sampling).

\section*{Acknowledgements}

This work was partially supported by Consejo Nacional de Investigaciones Cient\'{\i}ficas y T\'ecnicas (CONICET, Argentina) 
and Secretar\'{\i}a de Ciencia y Tecnolog\'{\i}a de la Universidad Nacional 
de C\'ordoba (SeCyT-UNC, Argentina).
This work used computational resources from CCAD\footnote{\url{https://ccad.unc.edu.ar/}}
 - Universidad Nacional de C\'ordoba,
which are part of SNCAD - MinCyT, Rep\'ublica Argentina.



\bibliographystyle{mnras}
\bibliography{references.bib}

\bsp
\label{lastpage}
\end{document}